\documentclass[aps,prd,superscriptaddress,nofootinbib,preprint]{revtex4}

\usepackage{ulem}
\usepackage{color}
\usepackage{graphicx}
\usepackage{multirow}
\usepackage{amsmath,amssymb,amsthm,amsxtra,overpic,bbm,bm,float
,epsfig}

\begin{document}

\title{Towards a complete reconstruction of supernova neutrino spectra in future large liquid-scintillator detectors}

\author{Hui-Ling Li}

\email{huiling@hepg.sdu.edu.cn}

\affiliation{School of Physics, Shandong University, Jinan 250100, China}

\affiliation{Institute of High Energy Physics, Chinese Academy of Sciences, Beijing 100049, China}

\author{Yu-Feng Li}

\email{liyufeng@ihep.ac.cn}

\affiliation{Institute of High Energy Physics, Chinese Academy of Sciences, Beijing 100049, China}

\affiliation{School of Physical Sciences, University of Chinese Academy of Sciences, Beijing 100049, China}

\author{Meng Wang}

\email{mwang@sdu.edu.cn}

\affiliation{School of Physics, Shandong University, Jinan 250100, China}

\author{Liang-Jian Wen}

\email{wenlj@ihep.ac.cn}

\affiliation{Institute of High Energy Physics, Chinese Academy of Sciences, Beijing 100049, China}

\author{Shun Zhou}

\email{zhoush@ihep.ac.cn}

\affiliation{Institute of High Energy Physics, Chinese Academy of Sciences, Beijing 100049, China}

\affiliation{School of Physical Sciences, University of Chinese Academy of Sciences, Beijing 100049, China}

\affiliation{Center for High Energy Physics, Peking University, Beijing 100871, China}

\begin{abstract}
In this paper, we show how to carry out a relatively more realistic and complete reconstruction of supernova neutrino spectra in the future large liquid-scintillator detectors, by implementing the method of singular value decomposition with a proper regularization. For a core-collapse supernova at a distance of $10~{\rm kpc}$ in the Milky Way, its $\overline{\nu}^{}_e$ spectrum can be precisely determined from the inverse beta-decay process $\overline{\nu}^{}_e + p \to e^+ + n$, for which a $20~{\rm kiloton}$ liquid-scintillator detector with the resolution similar to the Jiangmen Underground Neutrino Observatory (JUNO) may register more than 5000 events. We have to rely predominantly on the elastic neutrino-electron scattering $\nu + e^- \to \nu + e^-$ and the elastic neutrino-proton scattering $\nu + p \to \nu + p$ for the spectra of $\nu^{}_e$ and $\nu^{}_x$, where $\nu$ denotes collectively neutrinos and antineutrinos of all three flavors and $\nu^{}_x$ for $\nu^{}_\mu$ and $\nu^{}_\tau$ as well as their antiparticles. To demonstrate the validity of our approach, we also attempt to reconstruct the neutrino spectra by using the time-integrated neutrino data from the latest numerical simulations of delayed neutrino-driven supernova explosions.
\end{abstract}

\maketitle

\section{Introduction}

The precious observations of the neutrino burst from Supernova (SN) 1987A in the Large Magellanic Cloud by Kamiokande-II~\cite{Hirata:1987hu}, Irvine-Michigan-Brookhaven~\cite{Bionta:1987qt} and Baksan~\cite{Alekseev:1988gp} have essentially demonstrated that the delayed neutrino-driven explosion mechanism works well for a core-collapse supernova~\cite{Colgate:1966ax,Bethe:1990mw,Woosley:2002zz,Janka:2006fh, Janka:2017vcp}. Their experimental data are consistent with the inverse beta-decay $\overline{\nu}^{}_e + p \to e^+ + n$ (IBD) initiated by SN $\overline{\nu}^{}_e$ of an average energy $\langle E^{}_{\overline{\nu}^{}_e}\rangle \approx 12~{\rm MeV}$, and the total gravitational binding energy released mainly via neutrinos is about $3\times 10^{53}~{\rm erg}$~\cite{Jegerlehner:1996kx,Loredo:2001rx} for a duration of 10 seconds. However, it is impossible to get more useful and robust information on SN neutrino energy spectra from two dozens of neutrino events totally observed in all those three experiments. See, e.g., Ref.~\cite{Vissani:2014doa}, for an excellent overview on the statistical analysis of neutrino data from SN 1987A.

Thanks to recent tremendous progress in neutrino oscillation experiments, a few large water-Cherenkov (WC), liquid-scintillator (LS) and liquid-argon time projection chamber (LAr-TPC) neutrino detectors have already been in operation (e.g., Super-Kamiokande~\cite{Okumura:2016zyh}, Borexino~\cite{DAngelo:2016cua} and KamLAND~\cite{Tolich:2011zz}) or will be available in the near future (e.g., Hyper-Kamiokande~\cite{Yokoyama:2017mnt}, JUNO~\cite{An:2015jdp} and DUNE~\cite{Acciarri:2015uup}). In the WC and LS detectors, the IBD channel is dominant and most sensitive to SN $\overline{\nu}^{}_e$, while the charged-current reaction $\nu^{}_e + ~^{40}{\rm Ar} \to e^- + ~^{40}{\rm K}^*$ in the LAr-TPC detectors is crucially important for probing the early-time neutronization burst of $\nu^{}_e$~\cite{Ankowski:2016lab}. As was first pointed out in Ref.~\cite{Beacom:2002hs}, the elastic scattering of SN neutrinos of all flavors on the protons in the LS detectors can hopefully be implemented to extract the spectral information of $\nu^{}_\mu$ and $\nu^{}_\tau$ and their antiparticles (which will be collectively denoted as $\nu^{}_x$). Recently, this idea has been further studied in Ref.~\cite{Dasgupta:2011wg}, where it has been shown that the energy spectrum of SN $\nu^{}_x$ can be partially reconstructed from the elastic neutrino-proton scattering ($p$ES) events. Therefore, for a future galactic core-collapse SN, we will be able to achieve a high-statistics measurement of neutrino signals in multiple complementary channels.

In this paper, we study whether it is possible to accomplish a complete reconstruction of the energy spectra of SN neutrinos of all three distinct flavors, i.e., $\nu^{}_e$, $\overline{\nu}^{}_e$ and $\nu^{}_x$, from a single large LS detector, such as JUNO~\cite{An:2015jdp} under construction and the proposed LENA~\cite{Wurm:2011zn}. For the JUNO detector, which is designed to have 20 kiloton LS with a proton fraction about $12\%$, there will be about 5000 IBD events for a galactic SN at a distance of 10 kpc. At the same time, about 1500 $p$ES events will be registered above the visible-energy threshold of $0.2~{\rm MeV}$, while the number of elastic neutrino-electron ($e$ES) events is approximately 400~\cite{An:2015jdp, Lu:2016ipr}. For the LENA detector with the similar proton fraction and the energy resolution of $7\%$ at 1 MeV~\cite{Wurm:2011zn}, the corresponding statistics will be increased by a factor of about 2.5 due to the total target mass scaled upward by the same factor~\cite{Wurm:2011zn}. Apart from the target mass, the statistics will also be increased (or decreased) for a closer (or farther) SN. The motivation of a full reconstruction of SN neutrino energy spectra is two-fold. First, the formation of energy spectra of SN neutrinos is obviously associated with the production processes~\cite{Keil:2002in,Buras:2002wt,Janka:2017vlw}, so a direct extraction of neutrino energy spectra from future observations will be helpful in understanding the production of SN neutrinos and exploring the true explosion mechanism~\cite{Mirizzi:2015eza}. Second, besides the standard Mikheyev-Smirnov-Wolfenstein (MSW) matter effects~\cite{Wolfenstein:1977ue,Mikheev:1986gs}, significant flavor conversions of SN neutrinos induced by the coherent neutrino-neutrino scattering are likely to take place in the SN environment where a dense neutrino gas is present~\cite{Pantaleone:1992eq, Samuel:1993uw, Duan:2005cp, Duan:2006an, Hannestad:2006nj, Raffelt:2007yz}. See, e.g., Refs.~\cite{Duan:2009cd, Duan:2010bg, Chakraborty:2016yeg}, for recent reviews on collective neutrino oscillations and latest developments. To verify the spectral splits and swaps of SN neutrinos caused by their self-interactions, one has to determine SN neutrino spectra of individual flavors.

The present work differs from Ref.~\cite{Dasgupta:2011wg} at least in two aspects. First, the practically powerful approach of singular value decomposition (SVD) with a proper regularization is applied to unfold the energy spectra of $\nu^{}_e$, $\overline{\nu}^{}_e$ and $\nu^{}_x$ mainly from the $e$ES, IBD and $p$ES channels, respectively. The energy resolution of the future large LS detector is fully taken into account, while an approximate and analytical method is adopted only for $\nu^{}_x$ in Ref.~\cite{Dasgupta:2011wg}. In order to demonstrate the validity of the SVD approach, we also attempt to extract the SN neutrino spectra from the neutrino data produced by the numerical simulations of SN explosions. Second, for a single LS detector, we show that it is possible to accomplish a complete reconstruction of $\nu^{}_e$, $\overline{\nu}^{}_e$ and $\nu^{}_x$ spectra from the observations of a future galactic SN. Certainly, the combination of neutrino data from different types of detectors (e.g., Hyper-Kamiokande, JUNO and DUNE) running at the same time will further improve greatly the determination of SN neutrino spectra.

The remaining part of our paper is structured as follows. In Sec. II, we explain how the neutrino events in different channels are calculated for a typical JUNO-like detector. Then, the strategy to reconstruct SN neutrino spectra is outlined in Sec. III, where both the basic idea in the ideal case and the realistic SVD unfolding approach are briefly introduced. We show the numerical results of reconstruction in both cases of the parametrized neutrino spectra and the simulated ones in Sec. IV. Finally, we summarize in Sec. V our main conclusions.

\section{Supernova Neutrino Events}

The relevant reactions for SN neutrinos in the LS detectors have been studied in details in Refs.~\cite{An:2015jdp, Lu:2016ipr, Lujan-Peschard:2014lta}. Although the charged- and neutral-current interactions of SN neutrinos with the $^{12}{\rm C}$ and $^{13}{\rm C}$ nuclei are also available there, the corresponding numbers of neutrino events will be only sub-dominant~\cite{Lu:2016ipr}. For simplicity, we just focus on the dominant channels, i.e., IBD for $\overline{\nu}^{}_e$, $p$ES for $\nu^{}_x$ and $e$ES for $\nu^{}_e$, respectively.

In this work we take a $20~{\rm kiloton}$ LS detector with the resolution similar to JUNO as the working example to explore the reconstruction performance. The detector energy resolution is assumed to be $3\% / \sqrt{E^{}_{\rm o}/({\rm MeV})}$ with $E^{}_{\rm o}$ being the event observed energy, and the energy threshold is simply taken as $0.2~{\rm MeV}$ for both the $p$ES and $e$ES interaction channels. The calculations of neutrino events in these three channels are quite straightforward and will be briefly summarized below.

\subsection{SN Neutrino Spectra}

The mechanisms of SN neutrino production are distinct at three different stages of SN evolution, namely, the early-time neutronization burst, the accretion phase and the cooling phase. However, the differential neutrino fluences or time-integrated neutrino energy spectra can be perfectly described by the Keil-Raffelt-Janka (KRJ) parametrization~\cite{Keil:2002in}
\begin{eqnarray}
\frac{{\rm d}F^{}_\alpha}{{\rm d}E^{}_\alpha} = \frac{3.5\times 10^{13}}{{\rm cm}^{2}~{\rm MeV}} \cdot \frac{1}{4\pi D^2} \frac{\varepsilon^{}_\alpha}{\langle E^{}_\alpha \rangle} \frac{E^{\gamma^{}_\alpha}_\alpha}{\Gamma(1+\gamma^{}_\alpha)} \left(\frac{1+\gamma^{}_\alpha}{\langle E^{}_\alpha \rangle}\right)^{1+\gamma^{}_\alpha} \exp\left[ - (1+\gamma^{}_\alpha) \frac{E^{}_\alpha}{\langle E^{}_\alpha \rangle}\right] \; ,
%     (1)
\label{eq:KRS}
\end{eqnarray}
where the subscript $\alpha$ runs over three neutrino flavors $\nu^{}_e$, $\overline{\nu}^{}_e$ and $\nu^{}_x$, and the flavor-dependent total neutrino energy $\varepsilon^{}_\alpha$ is given in units of $5\times 10^{52}~{\rm erg}$. In addition, the distance $D$ to a galactic SN is normalized to a typical value of $10~{\rm kpc}$, the neutrino energy $E^{}_\alpha$ and average energy $\langle E^{}_\alpha \rangle$ are measured in MeV, and the spectral index $\gamma^{}_\alpha = 3$ will always be adopted in our calculations. Note that $\gamma^{}_\alpha = 2$ corresponds to the Maxwell-Boltzmann distribution, while $\gamma^{}_\alpha = 2.3$ to the Fermi-Dirac distribution with a zero chemical potential~\cite{Tamborra:2012ac}. In the assumption of energy equipartition, the total gravitational binding energy $E^{}_{\rm g} = 3\times 10^{53}~{\rm erg}$ released in a core-collapse SN within ten seconds is shared by $\nu^{}_e$, $\nu^{}_\mu$, $\nu^{}_\tau$ and their antiparticles, i.e., $\varepsilon^{}_\alpha \approx 5\times 10^{52}~{\rm erg}$. In our calculations, the nominal values of neutrino average energies will be set to $\langle E^{}_{\nu^{}_e}\rangle = 12~{\rm MeV}$, $\langle E^{}_{\overline{\nu}^{}_e} \rangle = 14~{\rm MeV}$ and $\langle E^{}_{\nu^{}_x} \rangle = 16~{\rm MeV}$. For the numerical simulations, the neutrino data at a given time provided by the Garching group~\cite{Hudepohl:2013} are fitted by the KRJ parametrization, and the total energy $\varepsilon^{}_\alpha$, the average energy $\langle E^{}_\alpha \rangle$ and the spectral index $\gamma^{}_\alpha$ are tabulated, so one has to integrate the time-dependent neutrino number fluxes over the emission time, instead of using Eq.~(\ref{eq:KRS}). In contrast, for the numerical simulation models from the Japan group~\cite{Nakazato:2013}, the neutrino data are provided for both time and energy distributions. In this case, we integrate over time to obtain the neutrino energy spectra for each simulation model.

\subsection{Three Dominant Channels}

\subsubsection{IBD Events}

Due to its large cross section and clear signal, the IBD is the primary reaction for SN neutrino detection in the LS detectors. For an average energy of $\langle E^{}_{\overline{\nu}^{}_e}\rangle = 14~{\rm MeV}$ and a galactic SN at $D = 10~{\rm kpc}$, a JUNO-like detector will register about 5000 IBD events, which are adequate for a precise determination of the SN $\overline{\nu}^{}_e$ spectrum with one percent errors~\cite{Lu:2016ipr} for both the total energy $\varepsilon^{}_{\overline{\nu}^{}_e}$ and the average energy $\langle E^{}_{\overline{\nu}^{}_e}\rangle$. The time coincidence of prompt and delayed signals from $e^+$ and $n$ in the final states, respectively, renders the IBD to be almost free of background. The IBD cross section can be calculated accurately, in particular for MeV neutrinos~\cite{Vogel:1999zy,Strumia:2003zx}. For illustration, we take the following simple formula
\begin{eqnarray}
\sigma^{}_{\rm IBD}(E^{}_{e^+}) = 9.52\times 10^{-44}~{\rm cm}^2 \left(\frac{E^{}_{e^+} p^{}_{e^+}}{{\rm MeV}^2}\right) \; ,
%     (2)
\label{eq:IBDxsec}
\end{eqnarray}
where the positron energy is given by $E^{}_{e^+} \approx E^{}_{\overline{\nu}^{}_e} - \Delta^{}_{np}$ with the neutron-proton mass difference $\Delta^{}_{np} \equiv m^{}_n - m^{}_p \approx 1.293~{\rm MeV}$, and the positron momentum is $p^{}_{e^+} = \sqrt{E^2_{e^+} - m^2_e}$. The energy threshold for incident neutrinos is $E^{\rm th}_{\overline{\nu}^{}_e} = \Delta^{}_{np} + m^{}_e \approx 1.804~{\rm MeV}$, while the visible energy in the detector is $E^{}_{\rm v} = E^{}_{e^+} + m^{}_e$ after the electron-positron annihilation. Therefore, the event spectrum can be figured out as
\begin{eqnarray}
\frac{{\rm d}N^{}_{\rm IBD}}{{\rm d}E^{}_{\rm o}} = N^{}_{\rm p} \int^\infty_{E^{\rm th}_{\overline{\nu}^{}_e}}  \sigma^{}_{\rm IBD}(E^{}_{\overline{\nu}^{}_e}) \cdot \frac{{{\rm d}F^{}_{\overline{\nu}^{}_e}}}{{\rm d}E^{}_{\overline{\nu}^{}_e}} \cdot {\cal G}(E^{}_{\rm o}; E^{}_{\rm v}, \delta^{}_E) ~{\rm d}E^{}_{\overline{\nu}^{}_e} \; ,
%     (3)
\label{eq:IBDevent}
\end{eqnarray}
where $E^{}_{\rm o}$ stands for the observed energy, and ${\cal G}(E^{}_{\rm o}; E^{}_{\rm v}, \delta^{}_E)$ for the Gaussian function of $E^{}_{\rm o}$ with $E^{}_{\rm v}$ and $\delta^{}_E = 3\% / \sqrt{E^{}_{\rm o} / ({\rm MeV})}$ being the expectation value and the standard deviation, respectively. Given the target mass of $20$ kiloton and a proton fraction of $12\%$ for the typical LS target, one find that the total number of protons is $N^{}_p \approx 1.4\times 10^{33}$.
%which is comparable to that at the Super-Kamiokande detector.

The integration of the event spectrum in Eq.~(\ref{eq:IBDevent}) over the observed energy $E^{}_{\rm o}$ gives rise to the total number of IBD events. Assuming that the gravitational binding energy takes the typical value of $E^{}_{\rm g} = 3\times 10^{53}~{\rm erg}$ for a core-collapse supernova and it is equally shared by neutrinos and antineutrinos of three flavors, which are emitted within a duration of ten seconds, one can obtain the number of IBD events around $5000$ for $\langle E^{}_{\overline{\nu}^{}_e} \rangle = 14~{\rm MeV}$ and $\gamma^{}_{\overline{\nu}^{}_e} = 3$~\cite{An:2015jdp}. In addition, the excellent energy resolution $\delta^{}_E = 3\% /\sqrt{E^{}_{\rm o}/({\rm MeV})}$ and simply linear relationship between the neutrino energy and the visible energy, namely, $E^{}_{\rm v} \approx E^{}_{\overline{\nu}^{}_e} - 0.782~{\rm MeV}$, assure an accurate enough determination of SN $\overline{\nu}^{}_e$ spectrum.

\subsubsection{$p$ES Events}

As an advantage of LS detectors, the $p$ES events can also be observed due to their low threshold of visible energies, if radioactive backgrounds and dark noises of photomultiplier tubes are well in control~\cite{An:2015jdp}. The general formula for the cross section of the $p$ES can be found in Ref.~\cite{Weinberg:1972tu}, and can be recast into the following form, which is particularly useful for supernova neutrinos~\cite{Beacom:2002hs}
\begin{eqnarray}
\frac{{\rm d}\sigma^{}_{\nu p}(E^{}_\nu)}{{\rm d}T^{}_{\rm p}} \approx 4.83\times 10^{-42}~{\rm cm}^2~{\rm MeV}^{-1} \left[1 + 466 \left(\frac{T^{}_{\rm p}}{\rm MeV}\right) \cdot \left(\frac{\rm MeV}{E^{}_\nu}\right)^2 \right] \; ,
%     (4)
\end{eqnarray}
where only the zeroth-order term in $E^{}_\nu/m^{}_p$ is retained, $T^{}_{\rm p}$ is the recoil energy of the final-state proton. The recoil energy will be significantly quenched in the LS, according to the Birks' law~\cite{Birks:1951boa}, such that only part of the kinetic energy yields the scintillation light, i.e.,
\begin{eqnarray}
T^\prime_{\rm p}(T^{}_{\rm p}) \approx \int^{T^{}_{\rm p}}_0  \frac{{\rm d}E}{1 + k^{}_{\rm B} \langle{\rm d}E/{\rm d}x\rangle} \; ,
%     (5)
\end{eqnarray}
where the Birks' constant $k^{}_{\rm B} \approx 0.01~{\rm cm}~{\rm MeV}^{-1}$ and $\langle {\rm d}E/{\rm d}x\rangle$ denotes the corresponding energy-loss rate of protons in the LS materials. Although the observation of the $p$ES events depends crucially on the radioactive backgrounds, such as the beta decays of the nuclear isotope $^{14}{\rm C}$ in the LS, and the dark noises, it is reasonable to assume that an energy threshold of $E^{\rm th}_{\rm o} = 0.2~{\rm MeV}$ for the single events is achievable in future LS detectors. In this case, the $p$ES event spectrum can be calculated as
\begin{eqnarray}
\frac{{\rm d}N^{}_{\nu p}}{{\rm d}T^\prime_{\rm p}} = N^{}_{\rm p} \sum_\alpha \frac{{\rm d}T^{}_{\rm p}}{{\rm d}T^\prime_{\rm p}} \int^{\infty}_{E^{\rm min}_\alpha} \frac{{{\rm d}F^{}_{\alpha}}}{{\rm d}E^{}_{\alpha}} \cdot \frac{{\rm d}\sigma^{}_{\nu p}(E^{}_\alpha)}{{\rm d}T^{}_{\rm p}} {\rm d}E^{}_\alpha \; ,
\label{eq:sim_nup}
%     (6)
\end{eqnarray}
which will be further convoluted with the energy resolution function ${\cal G}(E^{}_{\rm o}; T^\prime_{\rm p}, \delta^{}_E)$ to give the event spectrum with respect to the observed energy $E^{}_{\rm o}$.

Note that according to the kinematics of the $p$ES, to produce a proton with a recoil energy $T^{}_{\rm p}$, we require the neutrino energy to be above $E^{\rm min}_\alpha = (T^{}_{\rm p} m^{}_p/2)^{1/2}$. In consideration of the backgrounds, only the events with $E^{}_{\rm o} > E^{\rm th}_{\rm o}$ can be identified. The number of $p$ES events is significant due to the contributions of both $\nu^{}_\mu$ and $\nu^{}_\tau$ and their antiparticles, whose average energy is higher than those of $\nu^{}_e$ and $\overline{\nu}^{}_e$. It is easy to find that the total number of $p$ES events is about 1500 for $E^{\rm th}_{\rm o} = 0.2~{\rm MeV}$, while about 2000 for $E^{\rm th}_{\rm o} = 0.1~{\rm MeV}$. This observation indicates that the reduction of radioactive backgrounds and dark noises is crucially important for the reaction channels with singles.

\subsubsection{$e$ES Events}

The $e$ES is another very important channel to detect SN neutrinos in the sense that it is sensitive to the early-time $\nu^{}_e$ burst from neutronization, which has been found to be a robust feature of SN neutrinos from numerical simulations~\cite{Mirizzi:2015eza}. The cross sections of neutrinos scattering off electrons at the tree level are well known in the standard theory of electroweak interactions~\cite{tHooft:1971ucy,Marciano:2003eq}
\begin{eqnarray}
\frac{{\rm d}\sigma^{}_{\nu e}(E^{}_\nu)}{{\rm d}T^{}_{\rm e}} &=& \frac{2 m^{}_e G^2_{\rm F}}{\pi} \left[\epsilon^2_- + \epsilon^2_+ \left(1 - \frac{T^{}_e}{E^{}_\nu}\right)^2 - \epsilon^{}_- \epsilon^{}_+ \frac{m^{}_e T^{}_{\rm e}}{E^2_\nu}\right] \; ,
%     (7)
\end{eqnarray}
where the Fermi constant is $G^{}_{\rm F} = 1.166\times 10^{-5}~{\rm GeV}^{-2}$, and the kinetic energy of the final-state electron $T^{}_{\rm e} \equiv E^\prime_{\rm e} - m^{}_e$ is lying below $T^{\rm max}_{\rm e} = E^{}_\nu/[1 + m^{}_e/(2E^{}_\nu)]$. For electron neutrinos, the coefficients are given by $\epsilon^{}_- = -1/2 - \sin^2 \theta^{}_{\rm w}$ and $\epsilon^{}_+ = - \sin^2 \theta^{}_{\rm w}$; while $\epsilon^{}_- = 1/2 - \sin^2 \theta^{}_{\rm w}$ and $\epsilon^{}_+ = -\sin^2 \theta^{}_{\rm w}$ for muon and tau neutrinos. For the cross sections of antineutrinos, one has to perform the exchange of $\epsilon^{}_- \leftrightarrow \epsilon^{}_+$ in Eq.~(7).
%%%%%%%%%%%%%%%%%%%% Fig. 1%%%%%%%%%%%%%%%%%%%%%%%%%%%%%%%
\begin{figure}[!t]
\begin{center}
\begin{tabular}{l}
\includegraphics[width=0.55\textwidth]{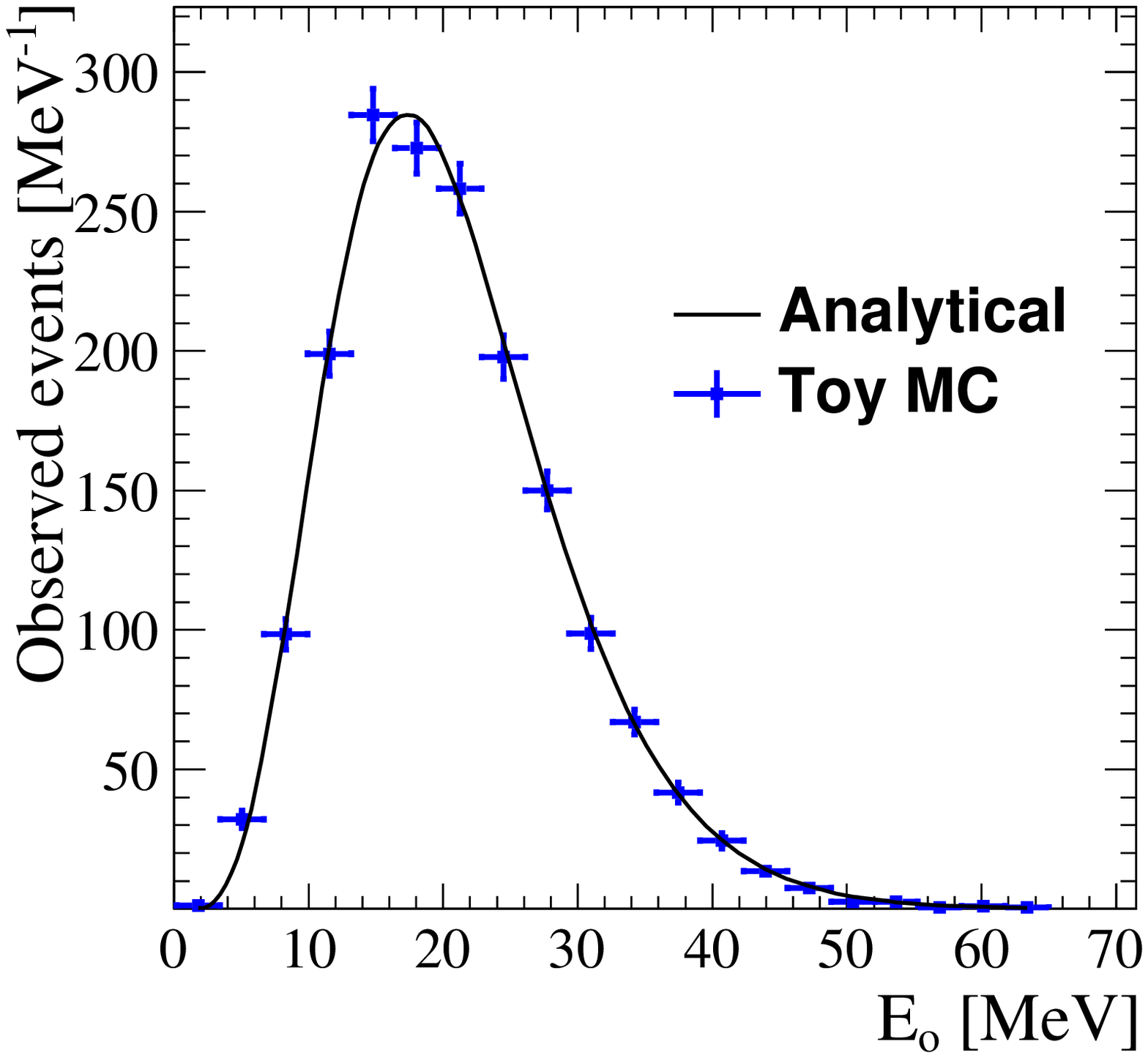}
\hspace{-1.5cm}
\includegraphics[width=0.55\textwidth]{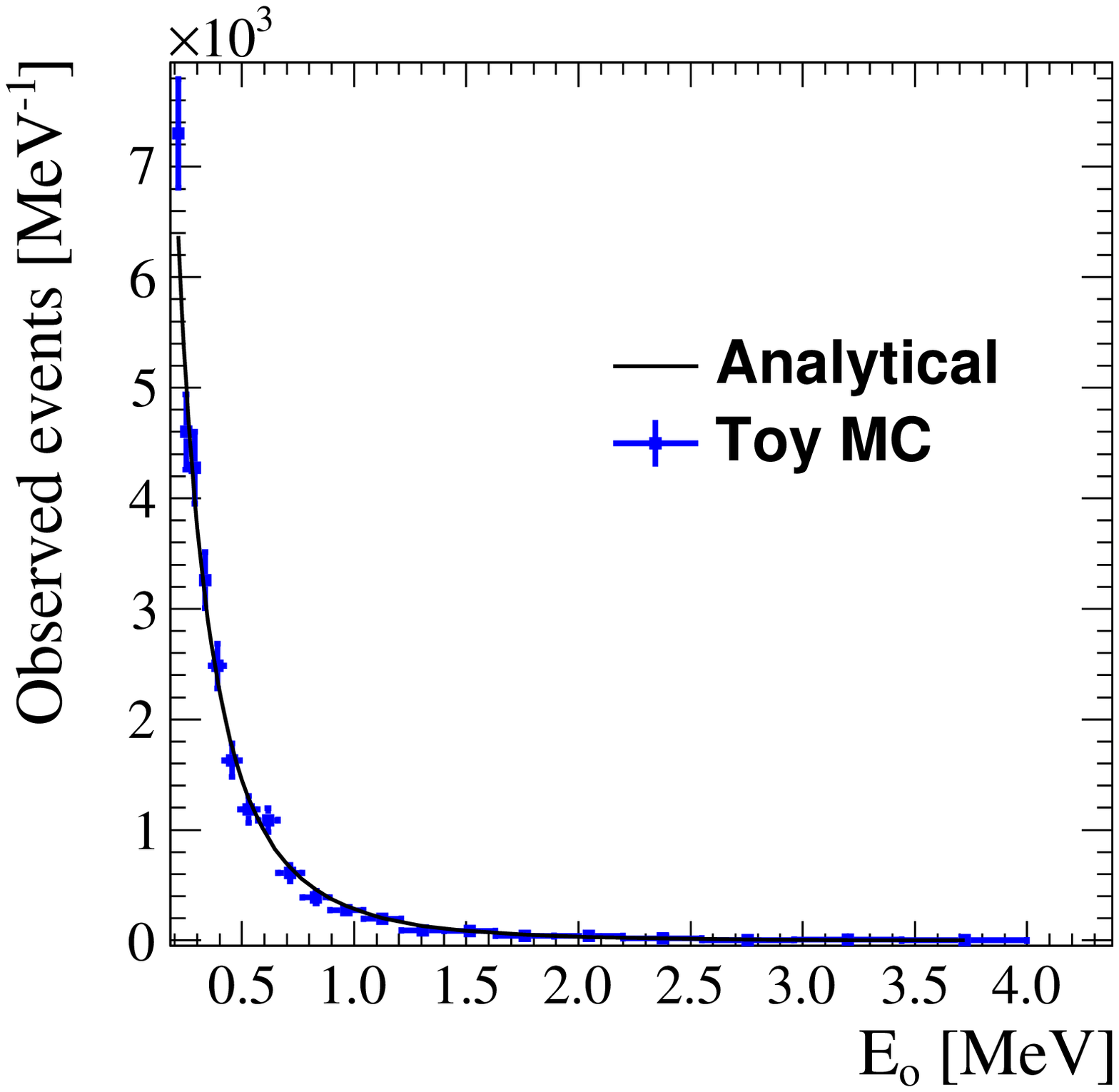}
\\
\includegraphics[width=0.55\textwidth]{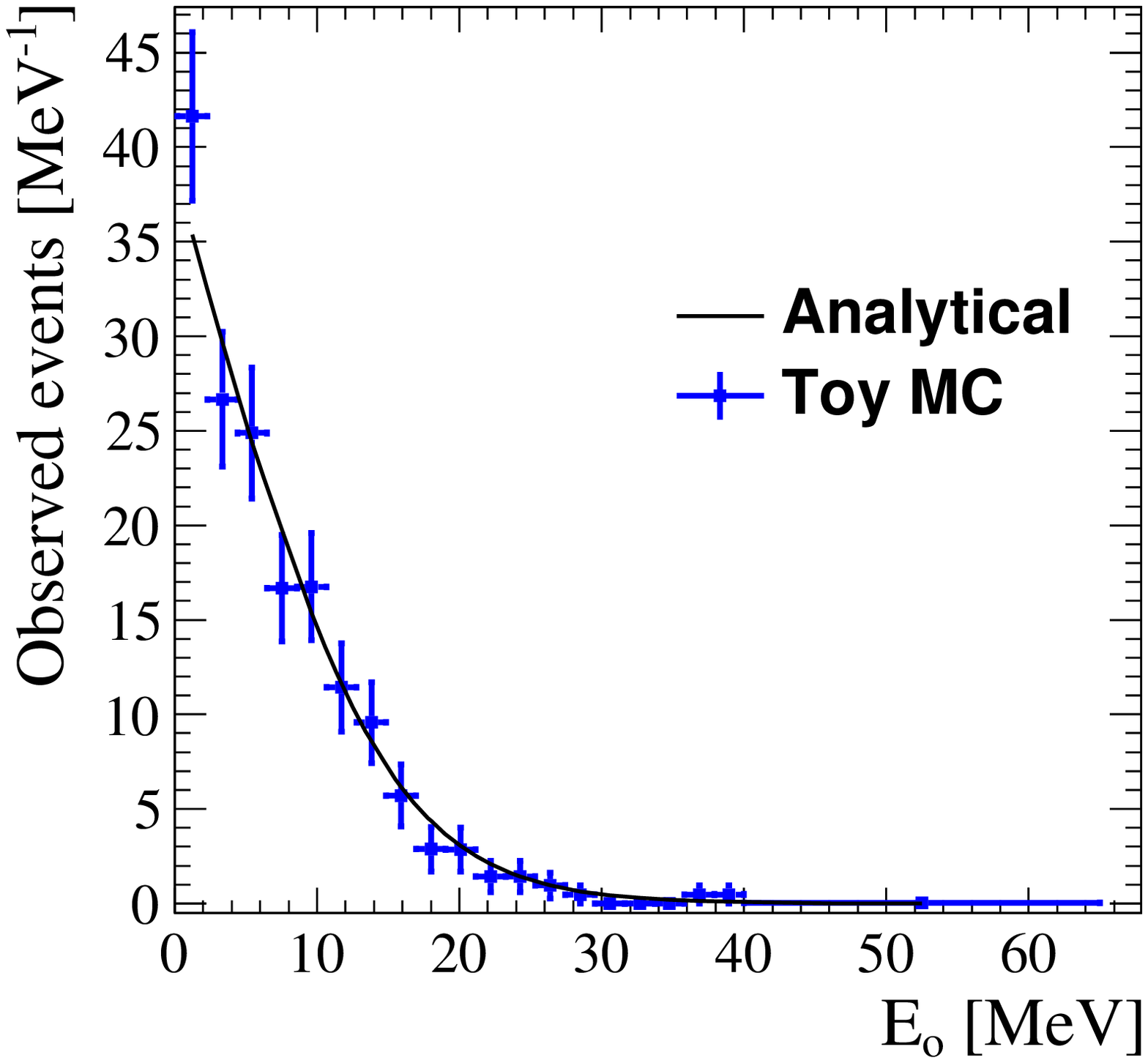}
\hspace{-1.5cm}
\includegraphics[width=0.55\textwidth]{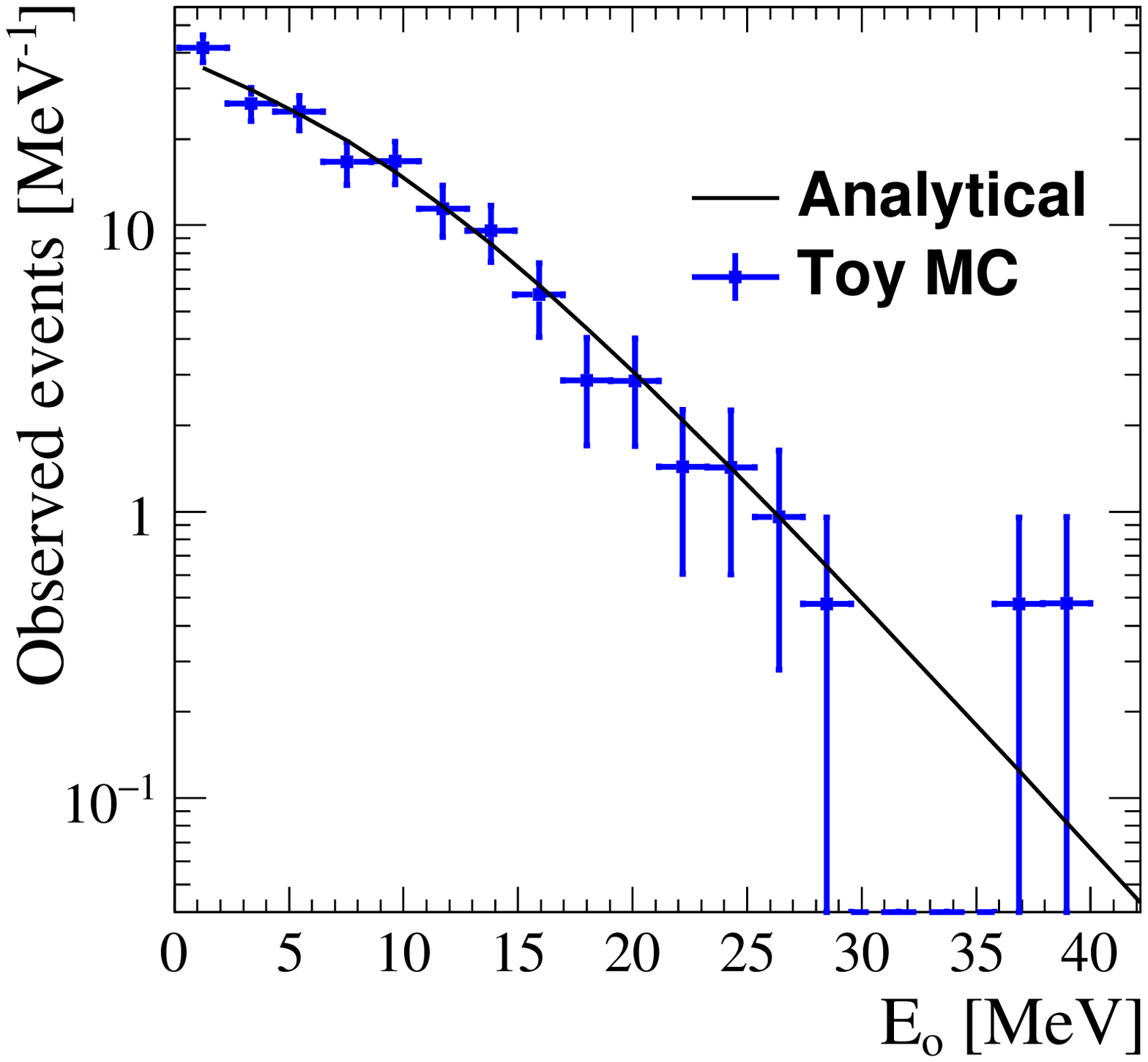}
\end{tabular}
\end{center}
\vspace{-0.5cm} \caption{The number of neutrino events at a JUNO-like LS detector for a galactic SN at 10 kpc, where the first row is for the IBD (left) and $p$ES (right) channels while the second row for the $e$ES channel. In the latter case, the results for both linear and logarithmic scales of the event number are shown in the left and right panel, respectively.
\label{fig:events}}
\end{figure}
%%%%%%%%%%%%%%%%%%%%%%%%%%%%%%%%%%%%%%%%%%%%%%%%%%%%%%%%%%
As the recoil energy of electrons will entirely be converted into the scintillation light, the observed event spectrum can be derived as follows
\begin{eqnarray}
\frac{{\rm d}N^{}_{\nu e}}{{\rm d}E^{}_{\rm o}} = N^{}_{\rm e} \sum_\alpha \int^\infty_0 {\rm d}T^{}_{\rm e} \cdot {\cal G}(E^{}_{\rm o}; T^{}_{\rm e}, \delta^{}_E) \int^\infty_{E^{\rm min}_\alpha} \frac{{{\rm d}F^{}_\alpha}}{{\rm d}E^{}_\alpha} \cdot \frac{{\rm d}\sigma^{}_{\nu e}(E^{}_\alpha)}{{\rm d}T^{}_e}~{\rm d}E^{}_\alpha \; ,
\label{eq:sim_nue}
%     (8)
\end{eqnarray}
where $N^{}_{\rm e}$ is the total number of electrons in the detector and the minimal neutrino energy $E^{\rm min}_\alpha \approx T^{}_{\rm e}/2 + \sqrt{T^{}_{\rm e}(T^{}_{\rm e} + 2m^{}_e)}/2$ required for an electron with a recoil energy of $T^{}_{\rm e}$ can be figured out from the kinematics. The contributions from neutrinos and antineutrinos of all three flavors are included, although the dominant one comes from electron neutrinos. Similar to the $p$ES events, the $e$ES events with energies below $E^{\rm th}_{\rm o} = 0.2~{\rm MeV}$ will be hidden in the radioactive backgrounds and dark noises.
%However, the low-energy $e$ES events are not dominant, so whether the threshold is $E^{\rm th}_{\rm o} = 0.2~{\rm MeV}$ or $0.1~{\rm MeV}$ is not very important.
%at {{a typical 20 kiloton LS detector with an energy resolution of $3\% / \sqrt{E^{}_{\rm o}/({\rm MeV})}$}}

For illustration, the IBD and $p$ES event spectra at a JUNO-like LS detector for a galactic SN at a distance of 10 kpc have been shown in the first row of Fig.~1. In the second row, the $e$ES events are displayed as well, where the number of events is given in both linear and logarithmic scales. In all these plots, the theoretical expectations of event numbers are denoted by black solid curves, while the toy Monte Carlo (MC) results by blue dots, together with the bin widths and statistical uncertainties being represented respectively by horizontal and vertical error bars. Note that the toy MC samples are randomly generated within the ROOT framework~\cite{Brun:1997pa} according to the event spectra in Eqs.~(\ref{eq:IBDevent}), (\ref{eq:sim_nup}) and (\ref{eq:sim_nue}), where an energy resolution of $3\% / \sqrt{E^{}_{\rm o}/({\rm MeV})}$ is assumed.

\section{Strategy for Reconstruction}

As demonstrated in Ref.~\cite{Lu:2016ipr}, the future large LS detectors could separately measure the average energies of SN $\nu^{}_e$, $\overline{\nu}^{}_e$ and $\nu^{}_x$ with reasonably good precisions, if the spectral indices are fixed at $\gamma^{}_\alpha = 3$ (for $\alpha = \nu^{}_e, \overline{\nu}^{}_e, \nu^{}_x$). However, without any presumed information, is it possible to reconstruct neutrino spectra of all three flavors directly from experimental data? In this work, we show that the answer is positive, and attempt to unfold all the SN neutrino spectra directly from simulated experimental neutrino data.

\subsection{The Ideal Case}
\label{ideal}

Ignoring the energy smearing, it is quite straightforward to obtain the SN $\overline{\nu}^{}_e$ spectrum from the IBD events. In practice, we divide the whole energy range of the observed IBD events into $N$ bins with the bin width $\Delta E^\prime_{i}$, namely, $[E^\prime_0, E^\prime_1], [E^\prime_1, E^\prime_2], \cdots, [E^\prime_{N -1}, E^\prime_N]$, and the number of events in each bin is denoted by $n^{}_i$. The event distribution can be represented by $n^{}_i/\Delta E^\prime_{i}$ at the central observed energy
%~\footnote{The simple approximation of using bin centers to define the average cross sections is only valid when the fluences are flat or the bin sizes are small enough.},
$\overline{E}^\prime_i = (E^\prime_i + E^\prime_{i-1})/2$ for $i = 1, 2, \cdots, N$.
According to the IBD event rate given in Eq.~(3) and the corresponding relationship $E^{}_i = E^\prime_i + 0.782~{\rm MeV}$ between the neutrino energy $E^{}_i$ and the observed energy $E^\prime_i$, we have
\begin{eqnarray}
f^{}_i = \left. \frac{{\rm d}F^{}_{\overline{\nu}^{}_e}}{{\rm d}E^{}_{\overline{\nu}^{}_e}}\right|^{}_{\overline{E}^{}_i} \simeq \frac{1}{N^{}_p \sigma(\overline{E}^{}_i)} \cdot \frac{n^{}_i}{\Delta E^\prime_{i}} \; ,
\label{IBD}
%     (9)
\end{eqnarray}
where $\overline{E}^{}_i = \overline{E}^\prime_i + 0.782~{\rm MeV}$ is the central value of neutrino energy in the bin $[E^{}_{i-1}, E^{}_i]$. Therefore, the initial $\overline{\nu}^{}_e$ spectrum is reconstructed as $f^{}_i$ at the energy $\overline{E}^{}_i$, where the same pattern of binning for the initial neutrino energy is assumed.

Next, we proceed with the reconstruction of the SN $\nu^{}_x$ spectrum from the $p$ES events.
%Since the average energy of $\nu^{}_x$ is larger than those of $\nu^{}_e$ and $\overline{\nu}^{}_e$, the $p$ES receives dominant contributions from $\nu^{}_x$. As a good approximation, we neglect the difference between the $p$ES cross sections of neutrinos and antineutrinos, and also the sub-leading contributions from $\nu^{}_e$ and $\overline{\nu}^{}_e$. In this case,
The distribution of $p$ES events can be denoted as $n^{}_i$ in the energy bin of $[T^\prime_{i-1}, T^\prime_i]$ with the central energy at $\overline{T}^\prime_i = (T^\prime_{i-1} + T^\prime_i)/2$ for $i = 1, 2, \cdots, N$.  Since the observed recoil energy $T^\prime_i$ of the final-state proton corresponds to $T^{}_i$ due to the quenching effects, one has to implement the Birks' law in Eq.~(5) to establish the relationship between $T^\prime_i$ and $T^{}_i$, or equivalently to convert the observed event spectrum into that with respect to $T^{}_i$. To be explicit, we have
\begin{eqnarray}
\frac{n^{}_i}{\Delta T^\prime_i} = \left.\frac{{\rm d}N}{{\rm d}T^\prime}\right|^{}_{\overline{T}^\prime_i} = \left.\frac{{\rm d}T}{{\rm d}T^\prime}\right|^{}_{\overline{T}^\prime_i} \cdot \left.\frac{{\rm d}N}{{\rm d}T}\right|^{}_{\overline{T}^{}_i} = \frac{n^{}_i}{\Delta T^{}_i} \; ,
%     (10)
\end{eqnarray}
where $\overline{T}^{}_i \equiv T^{}_{\rm p}(\overline{T}^\prime_i)$ with $T^{}_{\rm p}(T^\prime_{\rm p})$ being the inverse function of that defined in Eq.~(5) and $\Delta T^{}_i \equiv \Delta T^\prime_i \cdot \left.({\rm d}T/{\rm d}T^\prime)\right|^{}_{\overline{T}^\prime_i}$ is the bin width of the initial recoil energy. Furthermore, in order to produce the final-state proton with a recoil energy of $\overline{T}^{}_i$, we need an initial-state neutrino of energies above the minimum $E^i_{\rm min} = (\overline{T}^{}_i m^{}_p/2)^{1/2}$, and all the neutrinos with $E \geq E^i_{\rm min}$ can give rise to the $p$ES events in the bin at $\overline{T}^{}_i$.

Unlike in the IBD case, we shall now categorize the initial neutrino energy into $N$ different bins, i.e., $[E^{}_0, E^{}_1]$, $[E^{}_1, E^{}_2]$, $\cdots$, $[E^{}_{N- 1}, E^{}_{N}]$, where $E^{}_{i-1} = (T^{}_{i-1} m^{}_p/2)^{1/2}$ is fixed by the minimal neutrino energy to produce a recoil energy $T^{}_{i-1}$ for $i = 1, 2, \cdots, N$.
Since the average energy of $\nu^{}_x$ is larger than those of $\nu^{}_e$ and $\overline{\nu}^{}_e$, the $p$ES receives dominant contributions from $\nu^{}_x$. To illustrate the reconstruction method, we temporarily neglect the sub-leading contributions from $\nu^{}_e$ and $\overline{\nu}^{}_e$. All the contributions will be taken into account for the realistic case in the next section.
Defining $f^{}_j \equiv {\rm d}F^{}_{\nu^{}_x}/{\rm d}E^{}_{\nu^{}_x}|^{}_{\overline{E}^{}_j}$ as the fluence at $\overline{E}^{}_j \equiv (E^{}_j + E^{}_{j-1})/2$, one can establish the relationship
\begin{eqnarray}
\frac{n^{}_i}{\Delta T^\prime_i}  \simeq \sum_{j \in \{\overline{E}^{}_j \geq E^i_{\rm min}\}} 4 N^{}_{\rm p} \cdot \left.\frac{{\rm d}T}{{\rm d}T^\prime}\right|^{}_{\overline{T}^\prime_i} \cdot f^{}_j \Delta E^{}_j \cdot \left.\frac{{\rm d}\sigma^{}_{\nu p}(\overline{E}^{}_j)}{{\rm d}T^{}_{\rm p}}\right|^{}_{\overline{T}^{}_i} \equiv \sum_j f^{}_j K^{}_{ij} \; ,
\label{pES}
%     (11)
\end{eqnarray}
where the bin width $\Delta E^{}_j \equiv E^{}_j - E^{}_{j-1}$ and the matrix element $K^{}_{ij}$ can be identified without any ambiguity. Note that the summation is performed over the bin index $j$ corresponding to $\overline{E}^{}_j \geq E^i_{\rm min} = (\overline{T}^{}_i m^{}_p/2)^{1/2}$, so the matrix element $K^{}_{ij}$ vanishes for $i > j$.
To be explicit, we have
\begin{equation}
\begin{pmatrix}
\displaystyle \frac{n_{1}}{\Delta T_{1}^{\prime}} \\ \displaystyle \frac{n_{2}}{\Delta T_{2}^{\prime}} \\ \vdots \\ \displaystyle \frac{n_{N}}{\Delta T_{N}^{\prime}}
\end{pmatrix}
=\begin{pmatrix}
K_{11} & K_{12} &  \dots & K_{1N} \\
0 & K_{22} & \dots & K_{2N} \\
\vdots & \vdots & \ddots & \vdots \\
0 & 0 & \dots & K_{NN}
\end{pmatrix}
\begin{pmatrix}
f_{1}\\ f_{2} \\ \vdots \\ f_{N}
\end{pmatrix}
\; .
\end{equation}
One can choose the energy bins of the observed energy $T^\prime$ in such a way that the numbers of events in any bins are comparable in magnitude. Furthermore, the upper-triangular form of the matrix $K^{}_{ij}$ leads to an analytical inversion
\begin{eqnarray}
f^{}_j = \sum_i \left(K^{-1}\right)^{}_{ji} \cdot \frac{n^{}_i}{\Delta T^\prime_i} \; ,
%     (12)
\end{eqnarray}
where it is unnecessary to calculate $K^{-1}$ directly. The reason is simply that one can first obtain $f^{}_N = (n^{}_N/\Delta T^\prime_N)/K^{}_{NN}$ and then fix $f^{}_i$ iteratively for $i = N-1, N-2, \cdots, 2, 1$.

Finally, let us consider the $e$ES channel, for which the recoil energy of the final-state electron can be accurately measured and we assume that the possible backgrounds of other singles will be reduced by applying the approach of pulse shape discrimination. As before, the observed recoil energies of electrons can be divided into $N$ bins $[T^{}_0, T^{}_1]$, $[T^{}_1, T^{}_2]$, $\cdots$, $[T^{}_{N-1}, T^{}_N]$, where the recoil energies are identified as the observed energies without any quenching effects. In the $e$ES channel, since the cross section of $\nu_{e}$ is much larger than those of $\overline{\nu}^{}_e$ and ${\nu}^{}_x$, we take $\nu_{e}$ as an example to show the reconstruction method. Under this approximation, the $e$ES event spectrum can be readily calculated as
\begin{eqnarray}
\frac{n^{}_i}{\Delta T^{}_i} \simeq \sum_{j \in \{\overline{E}^{}_j \geq E^i_{\rm min}\}} N^{}_{\rm e} \cdot f^{}_j \Delta E^{}_j \cdot \left. \frac{{\rm d}\sigma^{}_{\nu e}(\overline{E}^{}_j)}{{\rm d}T^{}_{\rm e}}\right|^{}_{\overline{T}^{}_i} \; ,
%     (13)
\end{eqnarray}
where $n^{}_i$ stands for the event number in the observed energy bin $[T^{}_{i-1}, T^{}_i]$ with the width $\Delta T^{}_i = T^{}_i - T^{}_{i-1}$. In addition, the minimal neutrino energy $E^i_{\rm min}$ required to produce the recoil energy $\overline{T}^{}_i$ is given by $E^i_{\rm min} = \overline{T}^{}_i/2 + \sqrt{\overline{T}^{}_i(\overline{T}^{}_i + 2m^{}_e)}/2$, indicating that $E^{i}_{\rm min} \approx \overline{T}^{}_i$ for a relatively large recoil energy $\overline{T}^{}_i \gg 1~{\rm MeV}$. The initial neutrino energy bins $[E^{}_{j-1}, E^{}_j]$ for $j = 1, 2, \cdots, N$ can be chosen as $E^{}_{j-1} = E^{j-1}_{\rm min}$ such that all the energy bins with indices $j \geq i$ contribute to the event bin $i$. Therefore, one can immediately realize that the approach for the $p$ES channel can also be applied to the $e$ES channel. The number of $e$ES events, however, will be much smaller, leading to a worse precision on the extraction of the $\nu^{}_e$ spectrum.

\subsection{The Realistic Case}
\label{real}

%%%%%%%%%%%%%%%%%%%% Fig. 2%%%%%%%%%%%%%%%%%%%%%%%%%%%%%%%
\begin{figure}[!t]
\begin{center}
\begin{tabular}{c}
\vspace{-0.3cm}
\includegraphics[width=0.5\textwidth]{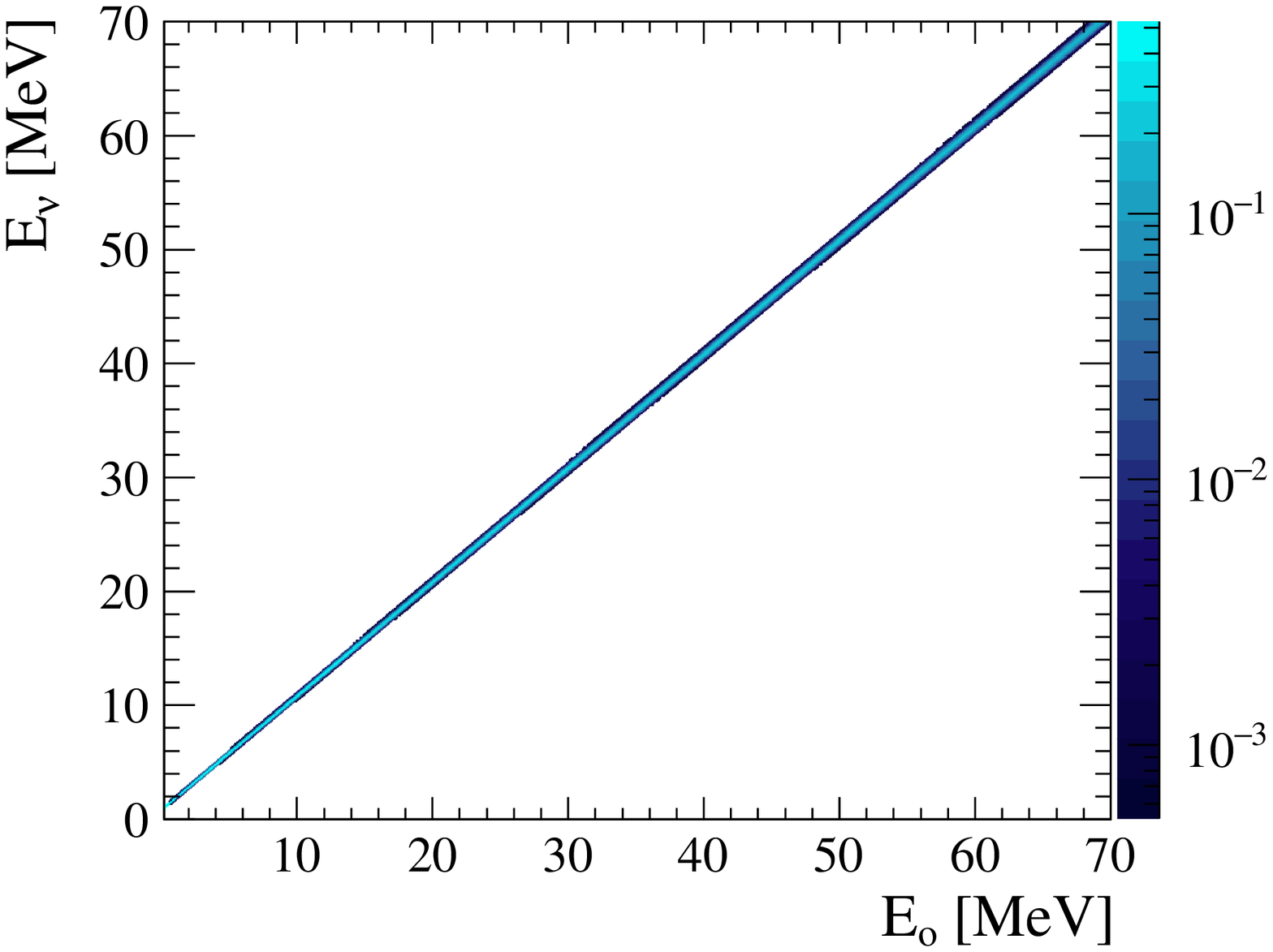}
\\
\vspace{-0.3cm}
\includegraphics[width=0.5\textwidth]{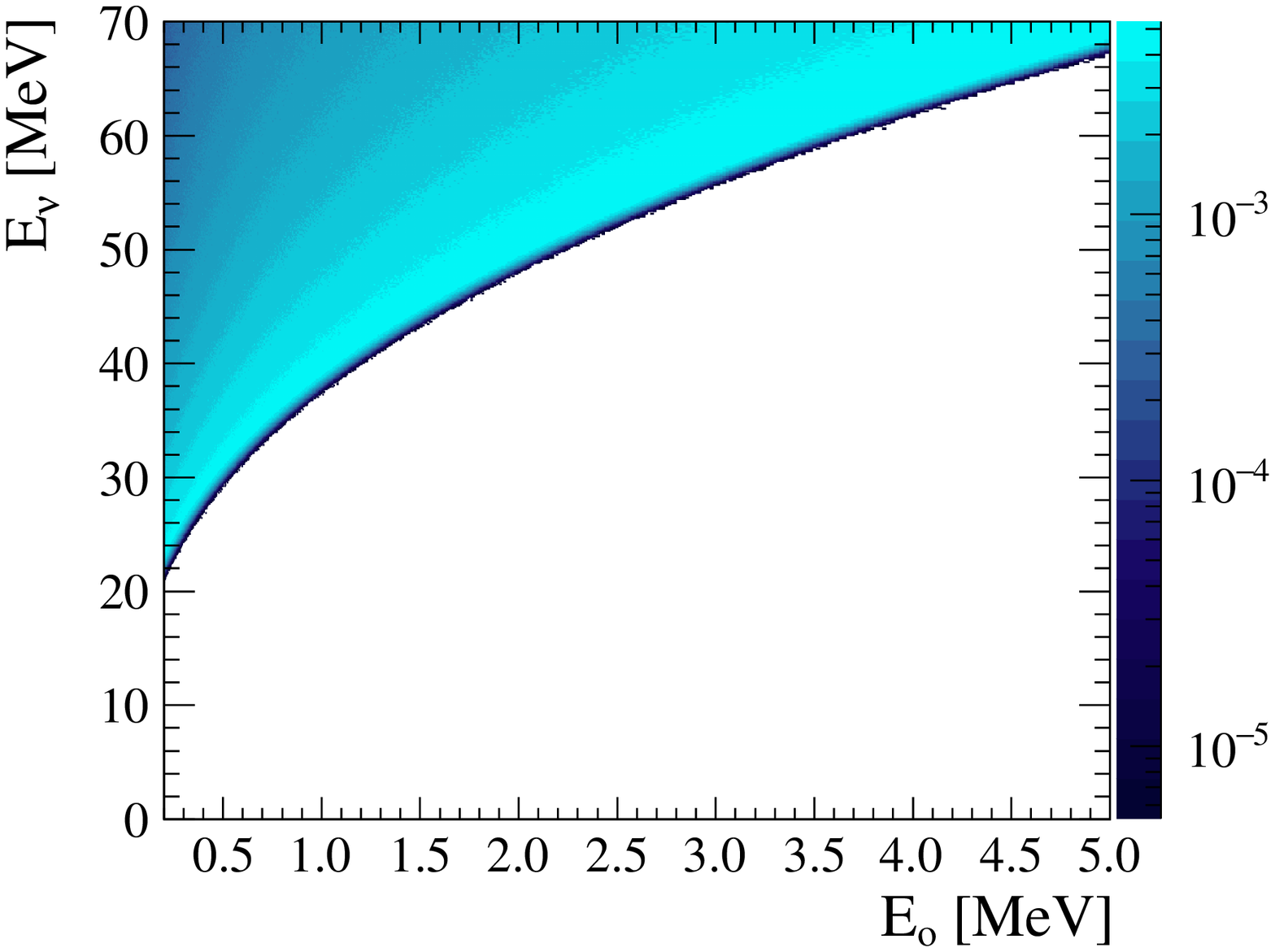}
\\
\vspace{-0.3cm}
\includegraphics[width=0.5\textwidth]{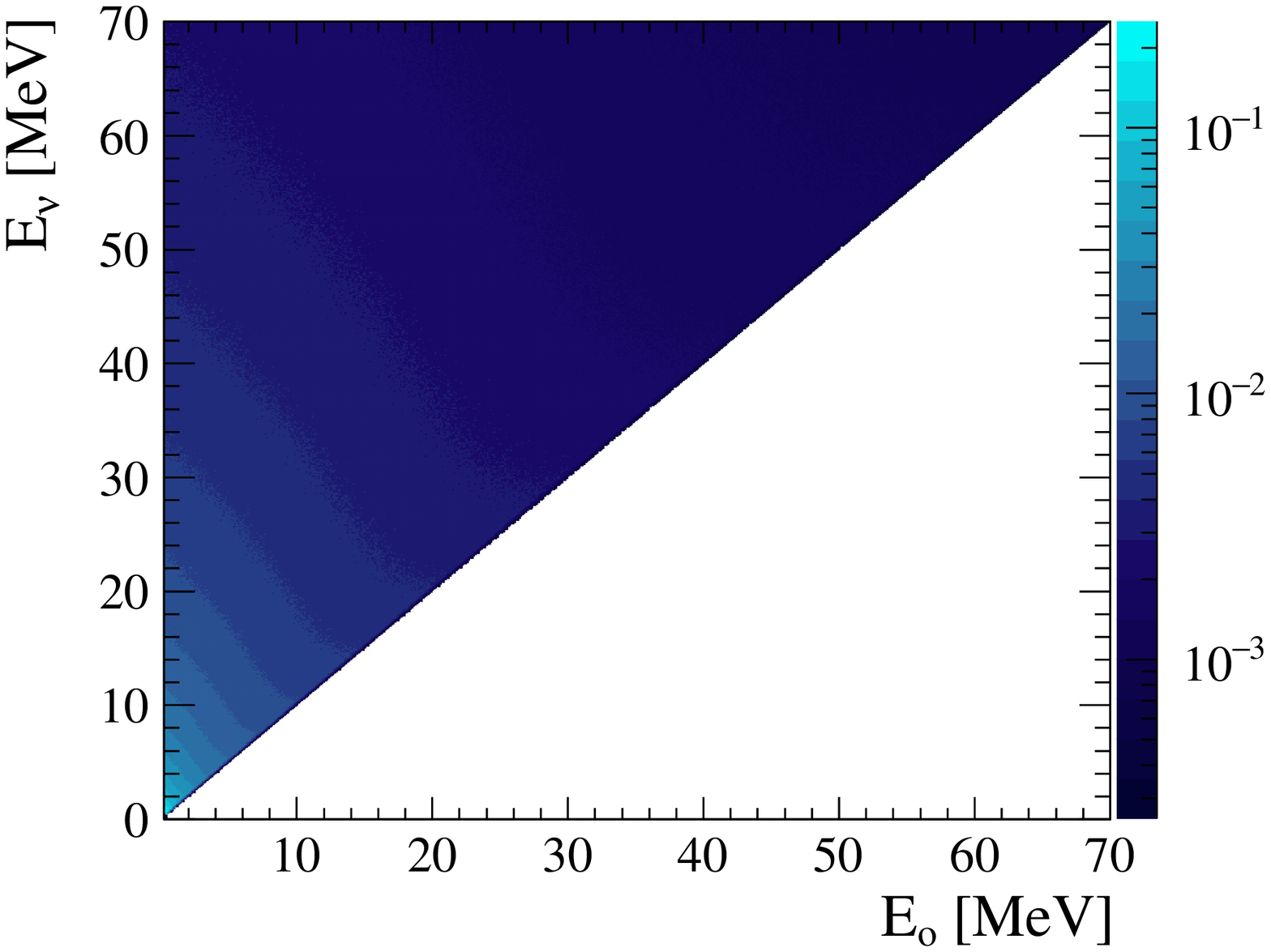}
\end{tabular}
\end{center}
\vspace{-0.5cm}
\caption{The detector response matrices for the IBD (upper), $p$ES (middle) and $e$ES (lower), where a large sample of events have been generated to get rid of any spurious statistical effects. The energy resolution of $3\% /\sqrt{E^{}_{\rm o}/({\rm MeV})}$ and the energy threshold of $E_{\rm o}^{\rm th}=0.2 ~{\rm MeV}$ have been taken for a JUNO-like LS detector.
\label{fig:response}}
\end{figure}
%%%%%%%%%%%%%%%%%%%%%%%%%%%%%%%%%%%%%%%%%%%%%%%%%%%%%%%%%%
In the ideal case, we have neglected the finite energy resolution and acceptance limitation of the detector. To incorporate them, we shall employ the unfolding technique to reconstruct the SN neutrino spectra. This technique is quite generic and frequently encountered in particle physics~\cite{Blobel:1984ku, Zech:2016gca} to extract the true distribution from experimental data. Here we follow the strategy proposed in Ref.~\cite{Hocker:1995kb}, where the SVD approach is clearly addressed and the necessity of an additional regularization is well explained. For completeness, we first briefly introduce the SVD method, and then apply it to the unfolding of SN neutrino spectra in the next section.

The spectral unfolding can be identified as a linear inverse problem of $\hat{A} x = b$, where $\hat{A}$ is an $m \times n$ response matrix characterizing the detector response effects (e.g., the energy resolution and the limited acceptance). Additionally, $x$ denotes an $n$-dimensional vector of the true distribution, while $b$ is an $m$-dimensional vector of the observed distribution. For the SN neutrino spectrum under consideration, $x$ is just the predicted neutrino event spectrum, which is roughly the convolution of the SN neutrino spectrum with the cross section, and $b$ is the observed event distribution. In fact, $\hat{A}$ can be calculated from a large sample of MC simulations, and for each bin of the true energy, the summation of each column of $\hat{A}$ (corresponding to the whole range of observed energies) is normalized to one.

To extract the true distribution, one can decompose the response matrix as $\hat{A} = U S V^{\rm T}$, where $U$ and $V$ are respectively the $m \times m$ and $n \times n$ orthogonal matrices while $S$ is an $m\times n$ diagonal matrix with only non-negative and non-increasing diagonal elements. Now the unfolding problem can be solved by using the generalized inverse matrix of $\hat{A}$. However, due to some small singular values in $S$ and the measurement errors in $b$, a direct extraction of $x$ with the exact inversion of the response matrix may lead to an unacceptable oscillatory solution. The standard way to suppress the spurious oscillating components is to introduce a regulator, which is actually an extra penalty term in the least-squares function of the corresponding inverse problem with a regularization parameter characterizing the weight of the penalty. The optimal value of the regularization parameter can be obtained by balancing the spurious oscillatory components and the bias caused by the regularization.

%For {{a typical LS detector with the target mass of $20$ kiloton}},
The response matrices of the IBD, $p$ES and $e$ES channels are depicted in Fig.~\ref{fig:response}, where 200 million neutrino events for each channel are simulated with flat distributions of the true neutrino spectra to avoid any informative priors on the inputs. The finite energy resolution of $3\% /\sqrt{E_{\rm o}/({\rm MeV})}$ and the LS quenching effect are taken into account in the simulations. In addition, the response matrices for the KRJ parametrization of SN neutrino spectra with three different average energies are also simulated and compared with the results of flat distributions. The differences between these two cases are found to be negligibly small because of the large statistics. In the upper panel of Fig.~\ref{fig:response} the response matrix for the IBD channel turns out to be rather simple. This can be well understood by noticing the relation $E^{}_{\overline{\nu}^{}_e} \approx E^{}_{\rm v} + 0.782~{\rm MeV}$ and the effect of energy resolution. On the other hand, the response matrices for $p$ES and $e$ES channels are more complicated in the realistic case as shown in the middle and lower panels of Fig.~\ref{fig:response}, where the kinematical relations between energies of the neutrino and proton/electron, the energy resolution and energy threshold, and the LS quenching effect are all properly implemented. In the calculations, we have assumed the threshold of observed energies to be $E_{\rm o}^{\rm th}=0.2 ~{\rm MeV}$, corresponding to the neutrino energy at about $20~{\rm MeV}$ in the $p$ES channel and $0.35~{\rm MeV}$ in the $e$ES channel.

After preparing the detector response matrices for the IBD, $p$ES and $e$ES channels, we can apply the SVD unfolding method to the reconstruction of the SN neutrino spectra. In the unfolding processes, the binning schemes for the observed spectra depend on the event statistics and are carefully treated in order to have comparable numbers of neutrino events in each observed energy bin. For the true energy spectra, we employ the equal-size binning schemes but combine the bins near the boundaries due to the limited statistics. The actual binning scheme for each unfolding process can be readily read out from the energy spectra which will be presented in the next section.
The practical realization of the SVD unfolding algorithm is based on the RooUnfold package~\cite{Adye:2011} and the choice for the optimal value of the regularization parameter can be found in Ref.~\cite{Hocker:1995kb}.

\section{Final Reconstruction Results}

% for {{a typical LS detector with the target mass of $20$ kiloton}}
In this section, we present the final results on the reconstruction of SN neutrino energy spectra. First of all, we simulate a large number of trials for SN neutrino data in the IBD, $p$ES and $e$ES channels by assuming the SN distance of $10~{\rm kpc}$ and the detector parameters discussed in Sec.~III. As we concentrate on the power of reconstruction, flavor conversions of SN neutrinos will temporarily be left aside. For the observed spectra in each trial, the SVD unfolding algorithm is applied to obtain the reconstructed neutrino spectra. To illustrate the dependence of the unfolding performance on the event statistics, we also simulate the neutrino data for the SN at the distances of $1.0~{\rm kpc}$ and $0.2~{\rm kpc}$ and reconstruct neutrino spectra in these two cases.
To explore the impact of regularization, we also compare the reconstruction results of neutrino spectra in the ideal case and those with the SVD unfolding algorithm. Moreover, we use the method of bin-to-bin separation to extract the original neutrino spectra of three flavors in their common energy range. Finally, we reconstruct the SN neutrino spectra by using the numerical SN models from both the Garching group~\cite{Hudepohl:2013} and the Japan group~\cite{Nakazato:2013}.

\subsection{Reconstructed Neutrino Energy Spectra}

%%%%%%%%%%%%%%%%%%%% Fig. 3%%%%%%%%%%%%%%%%%%%%%%%%%%%%%%%
\begin{figure}[!t]
\begin{center}
\begin{tabular}{c}
\hspace{-0.5cm}
\includegraphics[width=0.39\textwidth]{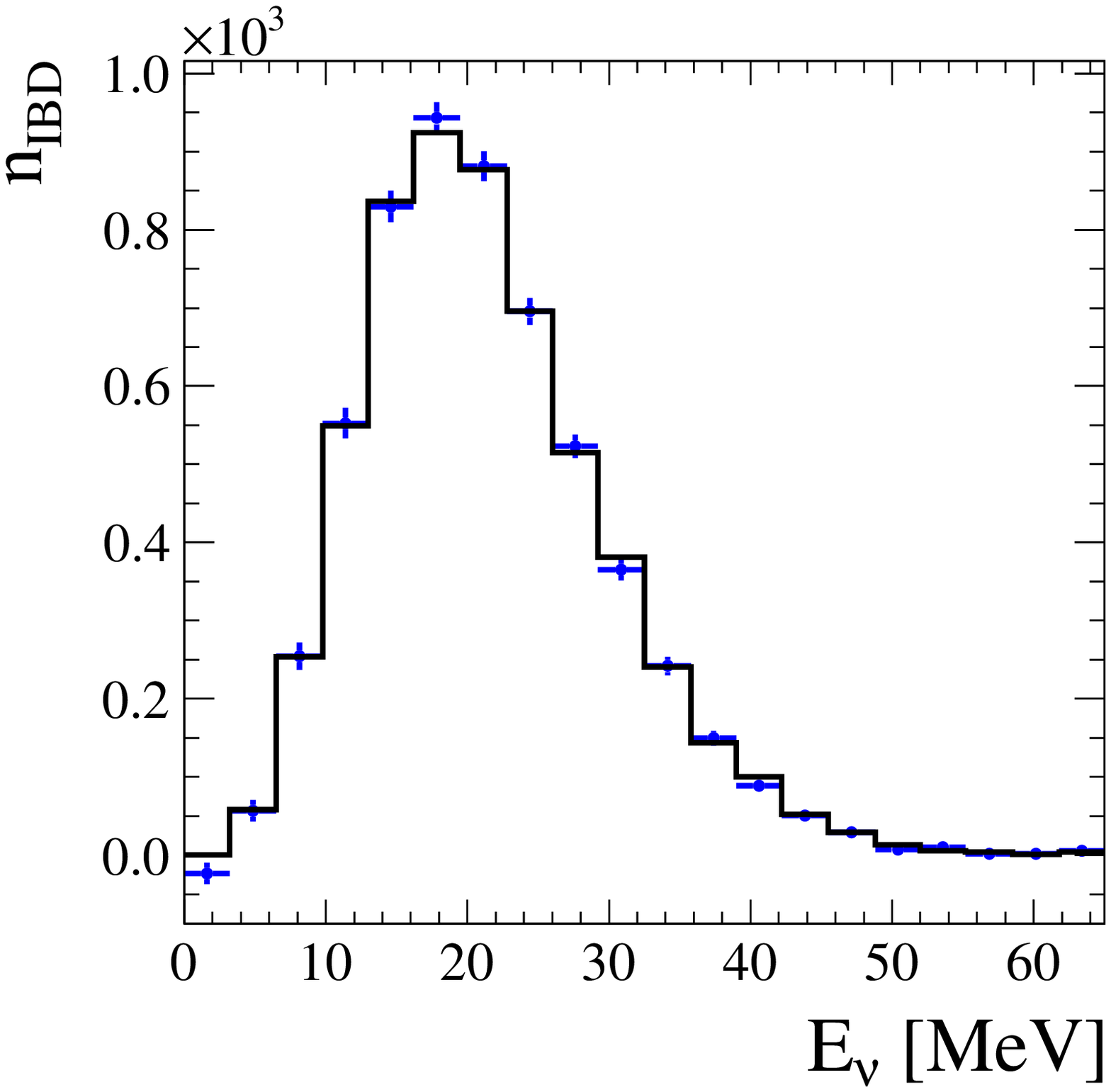}
\hspace{-1.1cm}
\includegraphics[width=0.39\textwidth]{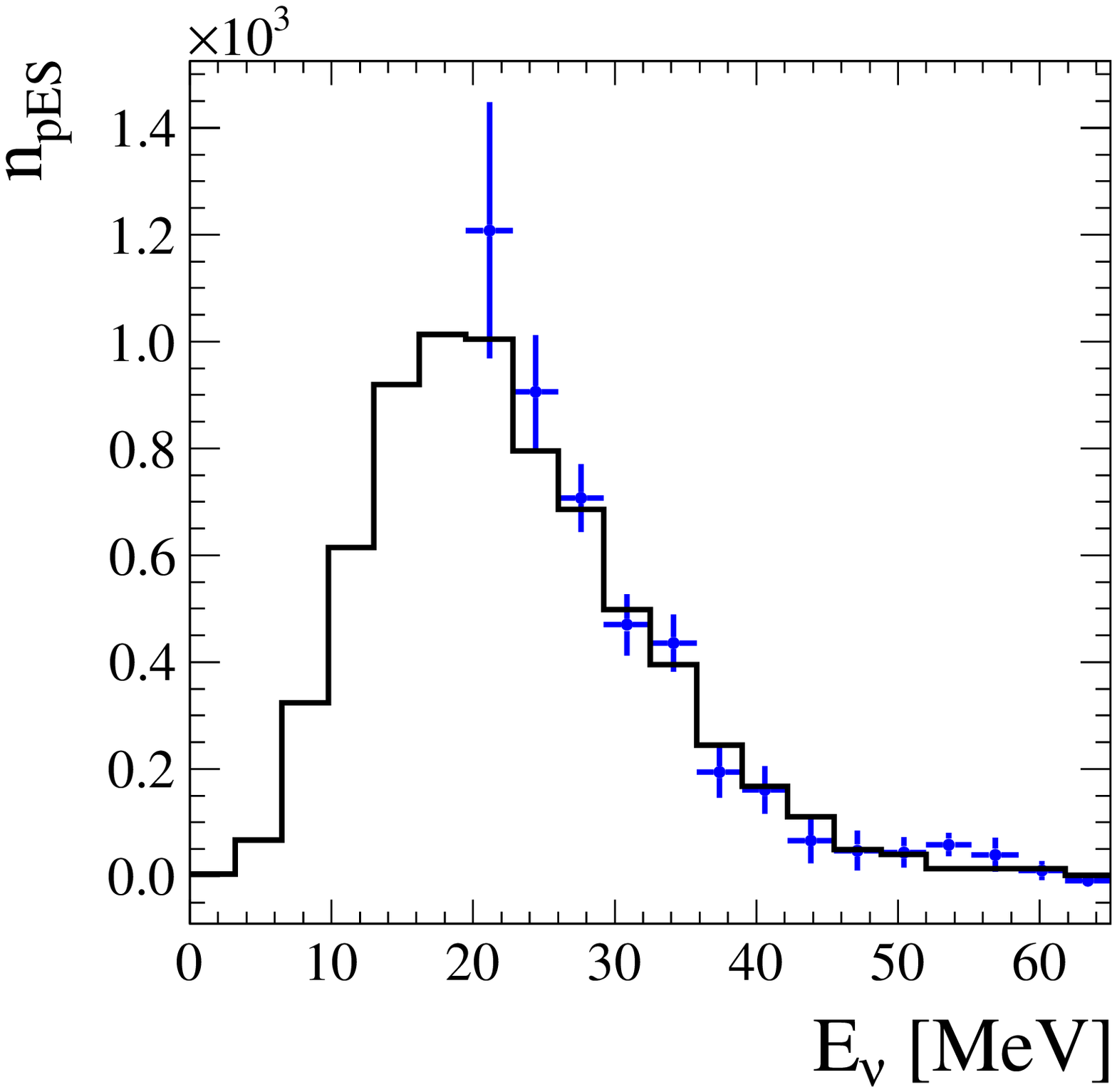}
\hspace{-1.1cm}
\includegraphics[width=0.39\textwidth]{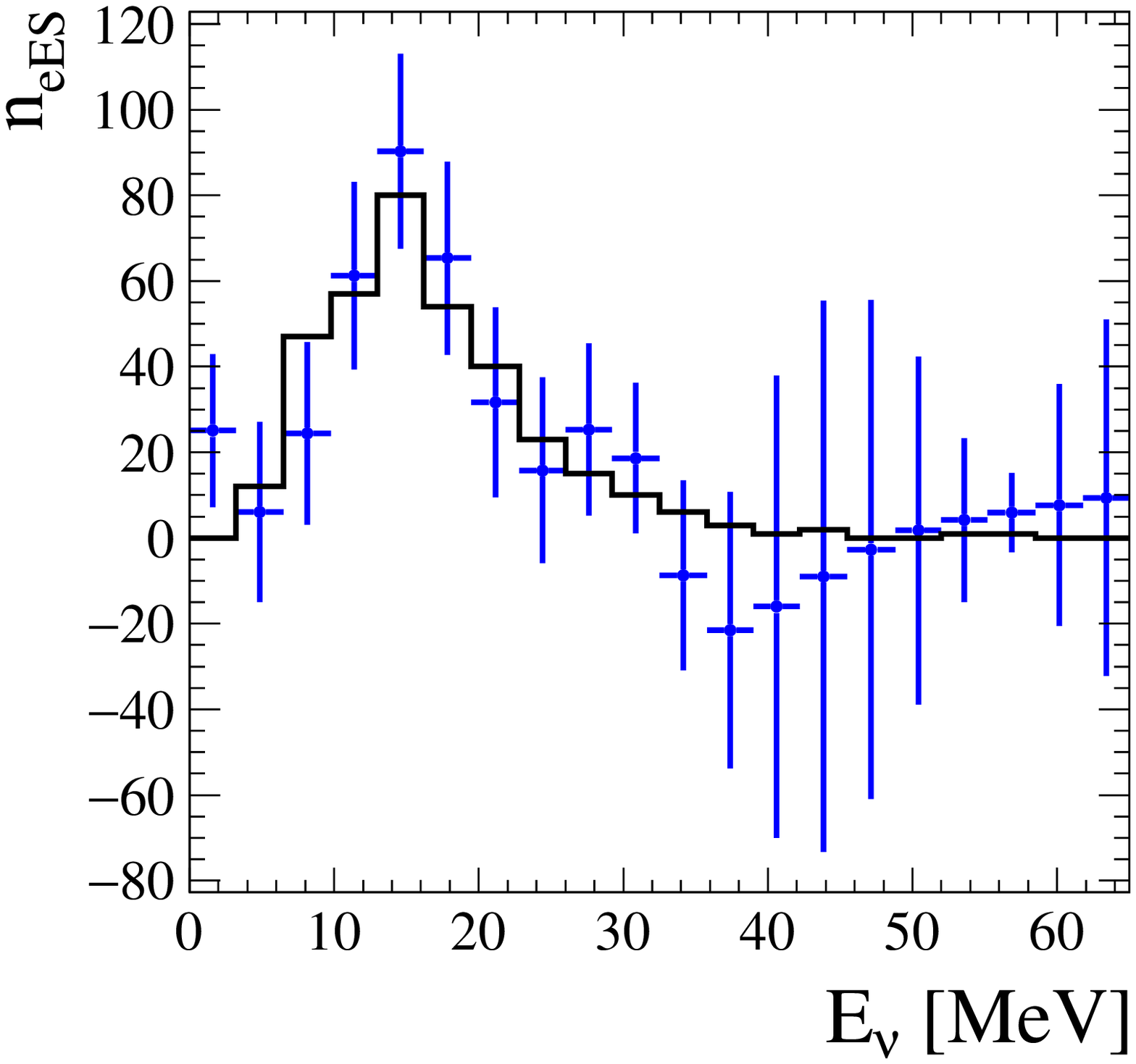}
\end{tabular}
\end{center}
\vspace{-0.6cm} \caption{Reconstructed SN neutrino spectra $n^{}_{\rm IBD}$,  $n^{}_{p\rm ES}$, and $n^{}_{e\rm ES}$ using the SVD unfolding technique for the IBD (left), $p$ES (middle) and $e$ES (right) channels. The simulated data are based on the KRJ parametrization of the SN neutrino fluences with $\langle E^{}_{\nu^{}_e}\rangle = 12~{\rm MeV}$, $\langle E^{}_{\overline{\nu}^{}_e} \rangle = 14~{\rm MeV}$ and $\langle E^{}_{\nu^{}_x} \rangle = 16~{\rm MeV}$ for a SN at $10~{\rm kpc}$. The black histograms are shown for the true SN neutrino energy spectra. The vertical error bars are the statistical uncertainties arising from those of the observed SN neutrino spectra while the horizontal ones show the bin widths.
\label{fig:svdUfdResult}}
\end{figure}
%%%%%%%%%%%%%%%%%%%%%%%%%%%%%%%%%%%%%%%%%%%%%%%%%%%%%%%%%%
The reconstructed neutrino energy spectra for the IBD, $p$ES and $e$ES channels, which are actually the SN neutrino fluences weighted with the corresponding cross sections, are shown in Fig.~\ref{fig:svdUfdResult}. The simulated data are for a SN at $10~{\rm kpc}$, and the KRJ parametrization of the SN neutrino fluences with $\langle E^{}_{\nu^{}_e}\rangle = 12~{\rm MeV}$, $\langle E^{}_{\overline{\nu}^{}_e} \rangle = 14~{\rm MeV}$ and $\langle E^{}_{\nu^{}_x} \rangle = 16~{\rm MeV}$ has been adopted. The true SN neutrino energy spectra are shown as black histograms. The vertical error bars are the statistical uncertainties arising from the those of the observed SN neutrino spectra, while the horizontal ones show the bin widths. From the plots one can observe that the reconstructed neutrino energy spectra are nicely in accordance with the true ones and their differences are within the statistical fluctuations which indicates that the biases from the regularization are significantly smaller than statistical uncertainties in the corresponding energy bins.

\begin{figure}%[!t]

\begin{center}
\begin{tabular}{cc}
\includegraphics*[width=0.45\textwidth]{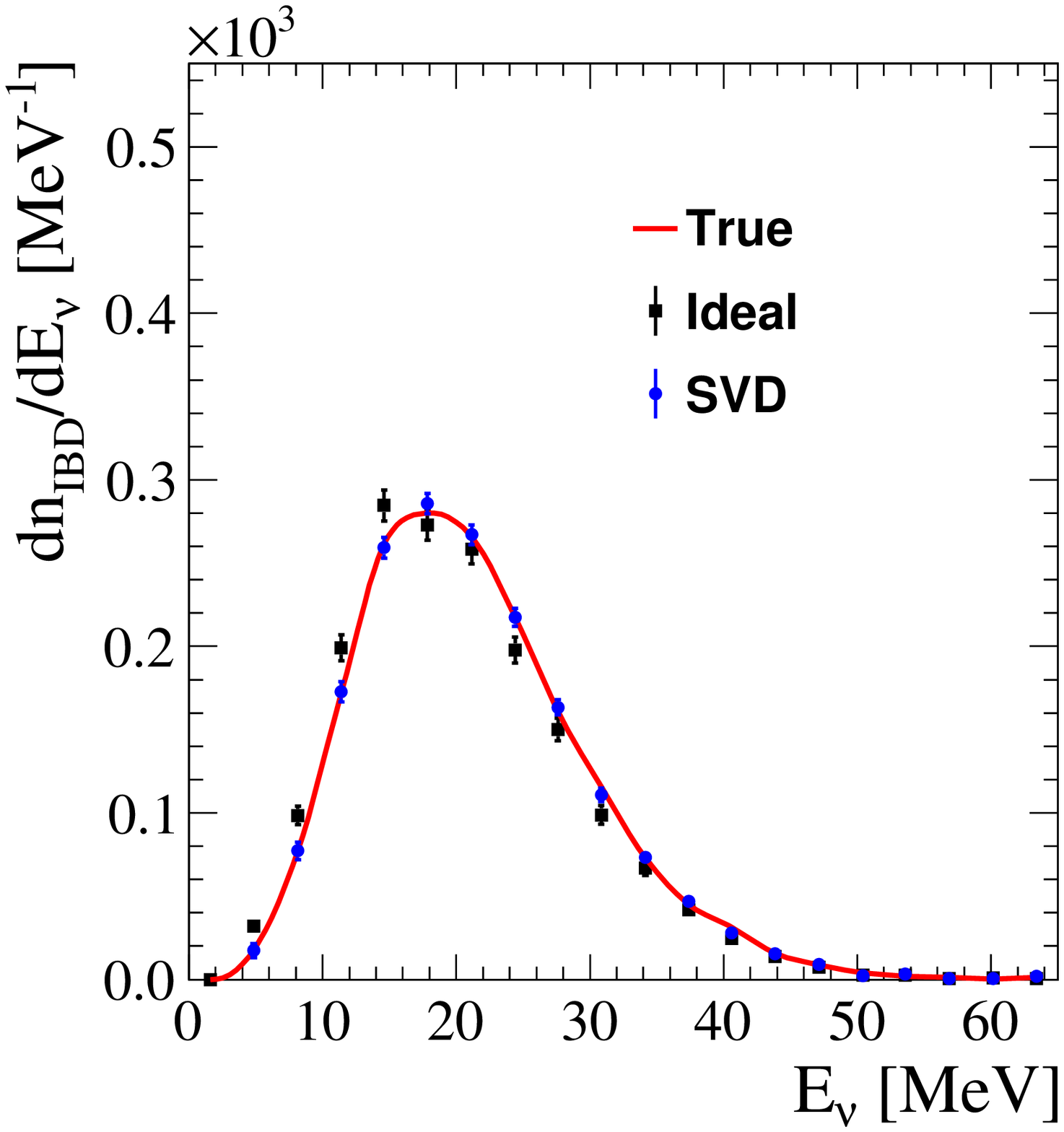}
&
\includegraphics*[width=0.45\textwidth]{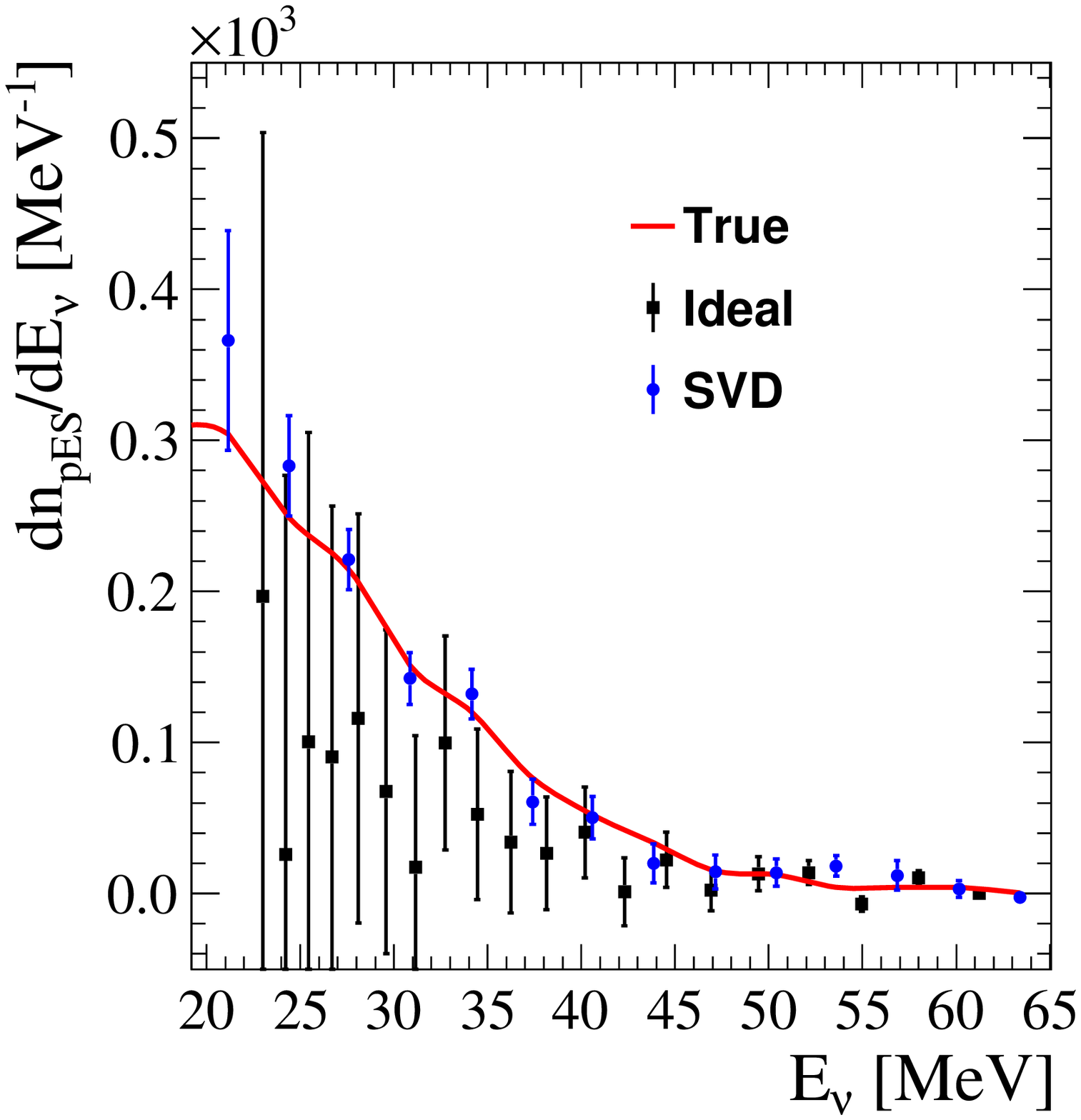}
%\includegraphics*[bb=3 20 480 500,width=0.45\textwidth]{IBD_IdealSVD.eps}
%&
%\includegraphics*[bb=22 18 488 450,width=0.45\textwidth]{pES_IdealSVD.eps}
\end{tabular}
\end{center}
%
%\begin{center}
%\begin{tabular}{c}
%%\hspace{-0.5cm}
%\includegraphics[width=0.45\textwidth]{IBD_IdealSVD.eps}
%%\hspace{-1.1cm}
%\includegraphics[width=0.45\textwidth]{pES_IdealSVD.eps}
%\end{tabular}
%\end{center}
\vspace{-0.6cm}
\caption{Reconstructed SN neutrino energy spectra by using the method in the ideal case and the SVD method for the IBD channel (left) and the $p$ES channel (right), where the same SN parameters as in Fig.~\ref{fig:svdUfdResult} have been adopted. The results of the ideal case are denoted by black square points and those using the SVD method by blue circle points. The error bars are statistical uncertainties propagated from the observed spectra.}
\label{fig:chaIdealSVD}
\end{figure}

To explore the impact of regularization, both the method in the ideal case and the SVD method are applied to the same observed spectra of the IBD and $p$ES channels as given in Fig.~\ref{fig:svdUfdResult}. The reconstructed neutrino spectra via these two methods, where the results of the ideal case are denoted by black square points and those using the SVD method by blue circle points, are shown in Fig.~\ref{fig:chaIdealSVD}. The error bars represent the statistical uncertainties propagated from the observed spectra. For the ideal case, the reconstructed spectra are calculated according to Eqs.~(\ref{IBD}) and (\ref{pES}) for the IBD and $p$ES channels, respectively. From such a comparison, one can easily observe the advantages of the SVD unfolding method, namely, more stable reconstructed results and lower statistical errors.
%%%%%%%%%%%%%%%%%%%% Fig. 5%%%%%%%%%%%%%%%%%%%%%%%%%%%%%%%
%%%%%%%%%%%%%%%%%%%% Fig. 5%%%%%%%%%%%%%%%%%%%%%%%%%%%%%%%

\begin{figure}%[!t]
\begin{center}
\begin{tabular}{c}
\vspace{-0.55cm}
\includegraphics[width=0.48\textwidth]{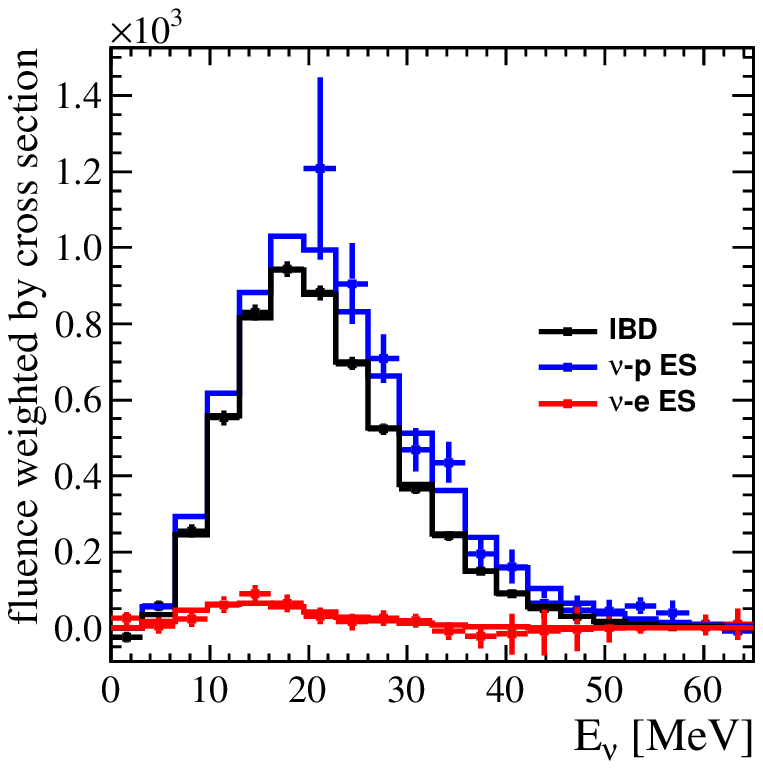}
\includegraphics[width=0.48\textwidth]{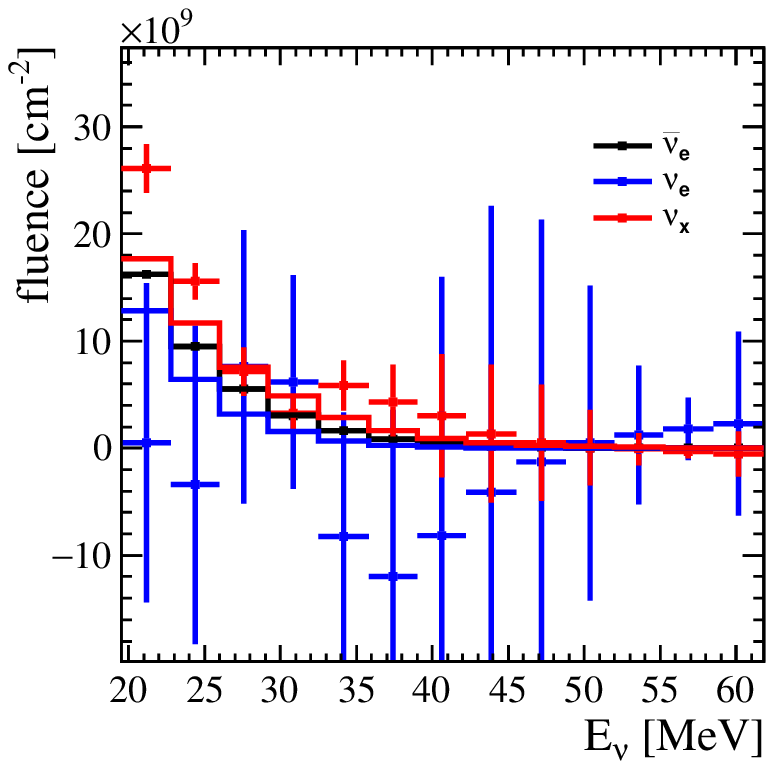}
\\
\vspace{-0.55cm}
\includegraphics[width=0.48\textwidth]{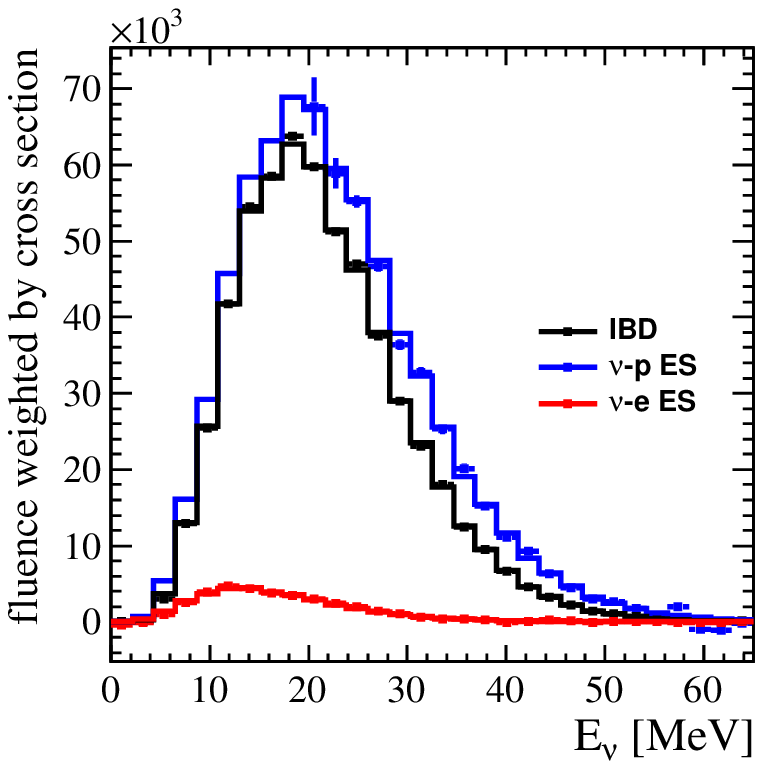}
\includegraphics[width=0.48\textwidth]{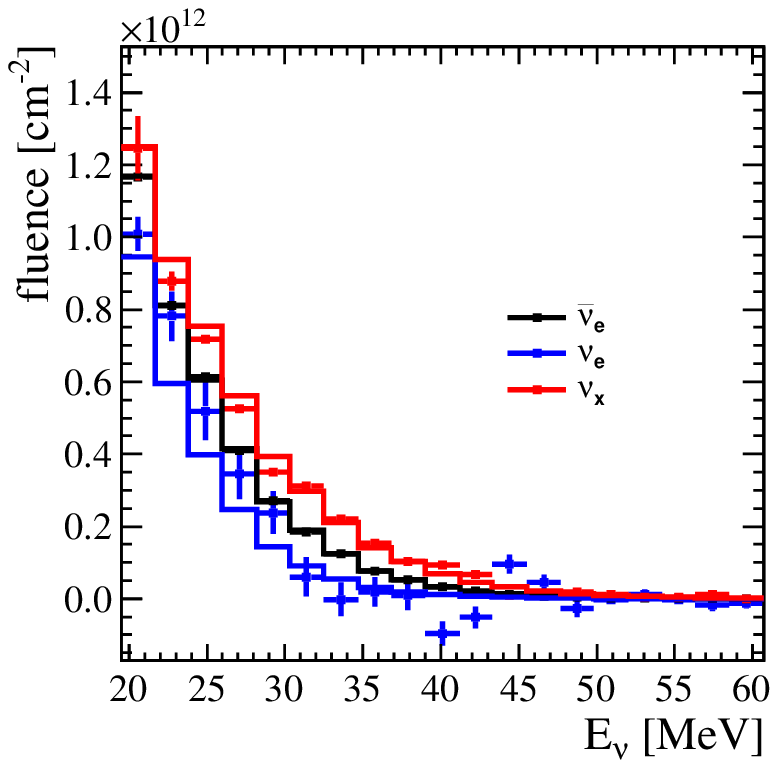}
\\
\vspace{-0.55cm}
\includegraphics[width=0.48\textwidth]{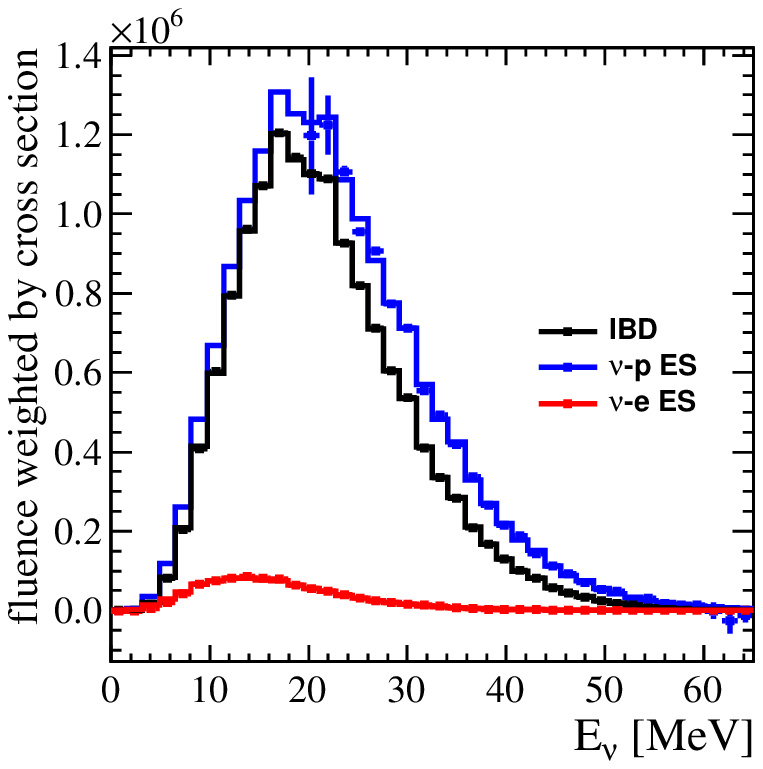}
\includegraphics[width=0.48\textwidth]{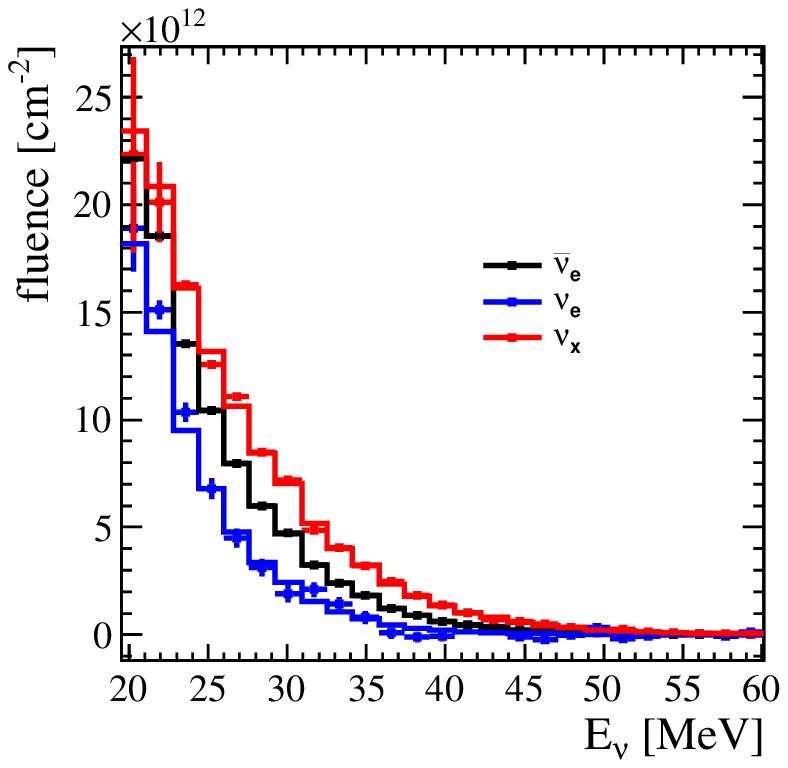}
\end{tabular}
\end{center}
\vspace{-0.3cm} \caption{The reconstructed SN neutrino energy spectra for the IBD, $p$ES and $e$ES channels (left) and for three flavor neutrinos (right) with the typical SN distances of 10 kpc, 1 kpc and 0.2 kpc for the upper, middle and lower panels respectively.
The histograms are shown for the true SN neutrino energy spectra.
The parameters of SN neutrino fluences are the same as in Fig.~\ref{fig:svdUfdResult}.
\label{fig:flavor}}
\end{figure}

In the left column of Fig.~\ref{fig:flavor}, the results of reconstructed neutrino spectra for the SN distances of $10~{\rm kpc}$, $1~{\rm kpc}$ and $0.2~{\rm kpc}$ are summarized. In each panel, the reconstructed spectra after unfolding for the IBD, $p$ES and $e$ES channels are represented by black, blue and red points, respectively.
The histograms are the corresponding true neutrino energy spectra. We can see that the reconstructed and true energy spectra well agree with each other within the statistical uncertainty. For the SN distance at 10 kpc, the best precision of the IBD, $p$ES and $e$ES channels can reach the levels of 2\%, 10\% and 50\%, respectively.
When the SN distances become closer and thus the statistics becomes larger, the precisions get better, indicating that the systematic bias of the unfolding method itself is negligible.
%%%%%%%%%%%%%%%%%%%% Fig. 5%%%%%%%%%%%%%%%%%%%%%%%%%%%%%%%
%%%%%%%%%%%%%%%%%%%%%%%%%%%%%%%%%%%%%%%%%%%%%%%%%%%%%%%%%%

\subsection{Original Neutrino Energy Spectra}

The reconstructed neutrino energy spectra from the IBD, $p$ES and $e$ES channels are essentially the original neutrino fluences weighted by cross sections. To further extract the original spectra within their common energy region (e.g., $E^{}_\nu \gtrsim 20~{\rm MeV}$), we choose the same binning scheme for the reconstructed neutrino spectra in all three channels and implement the bin-to-bin separation method, which can be clearly interpreted as the following equations
\begin{align}
n^{}_{\rm IBD}(E^{i}_{\nu})
=
\null & \null
N^{}_{\rm p}\sigma^{}_{\rm IBD}(E^{i}_{\nu}) F^{}_{\overline{\nu}^{}_{e}}(E^{i}_{\nu})\; ,
\label{3fIBD}
\\
n^{}_{p\rm ES}(E^{i}_{\nu})
=
\null & \null
N_{\rm p}
\left[\sigma^{}_{\nu^{}_{e}p}(E^{i}_{\nu})F^{}_{\nu^{}_{e}}(E^{i}_{\nu}) + \sigma^{}_{\overline{\nu}^{}_{e}p}(E^{i}_{\nu}) F^{}_{\overline{\nu}^{}_{e}}(E^{i}_{\nu}) +
4\sigma^{}_{\nu^{}_{x}p}(E^{i}_{\nu}) F^{}_{\nu^{}_{x}}(E^{i}_{\nu})\right]
\; ,
\label{3fpES}
\\
n^{}_{e\rm ES}(E^{i}_{\nu})
=
\null & \null
N^{}_{\rm e} \left[ \sigma^{}_{\nu^{}_{e}e}(E^{i}_{\nu}) F^{}_{\nu^{}_{e}}(E^{i}_{\nu}) + \sigma^{}_{\overline{\nu}^{}_{e}e}(E^{i}_{\nu}) F^{}_{\overline{\nu}^{}_{e}}(E^{i}_{\nu}) +
4 \sigma^{}_{\nu^{}_{x}e}(E^{i}_{\nu}) F^{}_{\nu^{}_{x}}(E^{i}_{\nu})\right]
\; ,
\label{3feES}
\end{align}
where $n^{}_{\rm IBD}(E^i_\nu)$, $n^{}_{p\rm ES}(E^i_\nu)$ and $n^{}_{e\rm ES}(E^i_\nu)$ stand for the reconstructed distributions in the neutrino energy bin $E^i_\nu$ for the IBD, $p$ES and $e$ES channels, and $F^{}_{\alpha}$ (for $\alpha = \nu^{}_{e}, \overline{\nu}^{}_{e} , \nu^{}_{x})$ are the original SN neutrino fluences.
Here $N^{}_{\rm p}$ and $N^{}_{\rm e}$ are the numbers of free protons and electrons in the LS target, while $\sigma^{}_{\rm IBD}$, $ \sigma^{}_{\alpha p}$, $ \sigma^{}_{\alpha e}$ are the cross sections of the IBD, $p$ES and $e$ES reactions, respectively. By solving the above three equations for each neutrino energy bin $E^i_\nu$, we achieve a complete reconstruction of the SN $\nu^{}_{e}$, $\overline{\nu}^{}_{e}$ and $\nu^{}_{x}$ energy spectra, as shown in the right column of Fig.~\ref{fig:flavor} for the SN distances of 10 kpc, 1 kpc and 0.2 kpc.

From the upper-right plot of Fig.~\ref{fig:flavor}, where the reconstruction has been performed in the case of a SN distance of 10 kpc, it is clear that the $\bar{\nu}_{e}$ energy spectra from the IBD channel can be determined with a precision better than $2\%$ in the vicinity of $E^{}_\nu = 20~{\rm MeV}$. The high-energy tail of the $\nu_{x}$ spectrum starting from 20 MeV are also well reconstructed, where the statistical uncertainty increases from $10\%$ at $E^{}_\nu = 20~{\rm MeV}$ to $30\%$ at $E^{}_\nu = 30~{\rm MeV}$. However, for just one trial of the SN $\nu^{}_{e}$ events at the distance of 10 kpc, it is difficult to provide an accurate reconstruction of the $\nu^{}_{e}$ spectrum because of the large statistical fluctuation, which can be at the level of 100\%.
On the other hand, with a much larger statistics when the SN distance is either 1 kpc or 0.2 kpc, one can observe an excellent reconstruction performance for all three neutrino flavors. This demonstrates the powerful capability of the SVD unfolding method. It is worthwhile to point out that SN $\nu^{}_e$ can also be observed in the LS detectors via the charged-current ${\nu}_{e}$-$^{12}$C interaction~\cite{Lu:2016ipr}, thus the incorporation of the ${\nu}_{e}$-$^{12}$C channel will improve the precision of reconstruction.

\subsection{Reconstruction with Numerical SN Models}
%%%%%%%%%%%%%%%%%%%% Fig. 5%%%%%%%%%%%%%%%%%%%%%%%%%%%%%%%
\begin{figure}%[!t]
\begin{center}
\begin{tabular}{l}
\vspace{-0.4cm}
\includegraphics[width=0.54\textwidth]{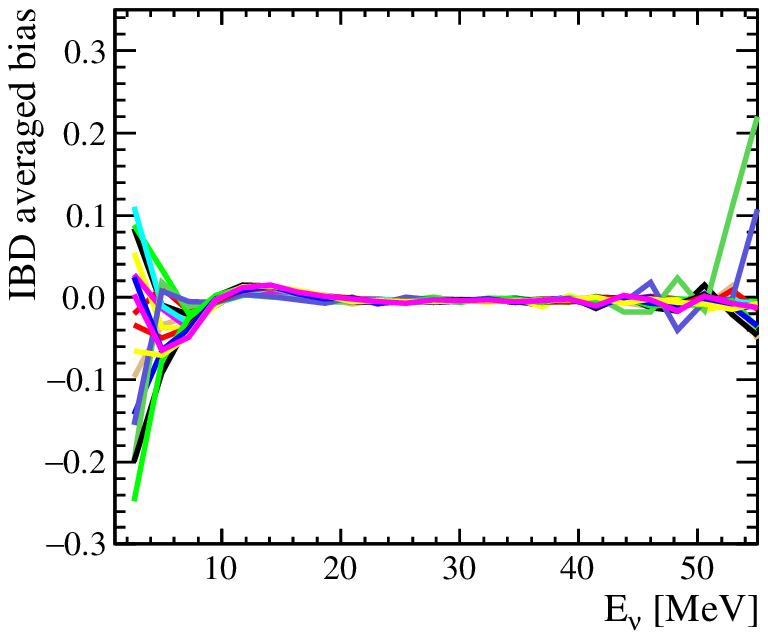}
\hspace{-0.8cm}
\includegraphics[width=0.54\textwidth]{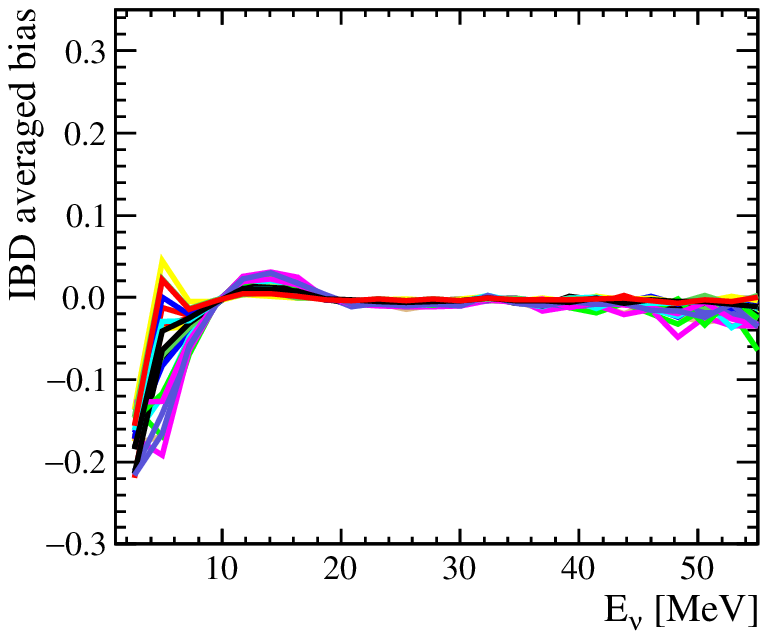}
\\
\vspace{-0.4cm}
\includegraphics[width=0.54\textwidth]{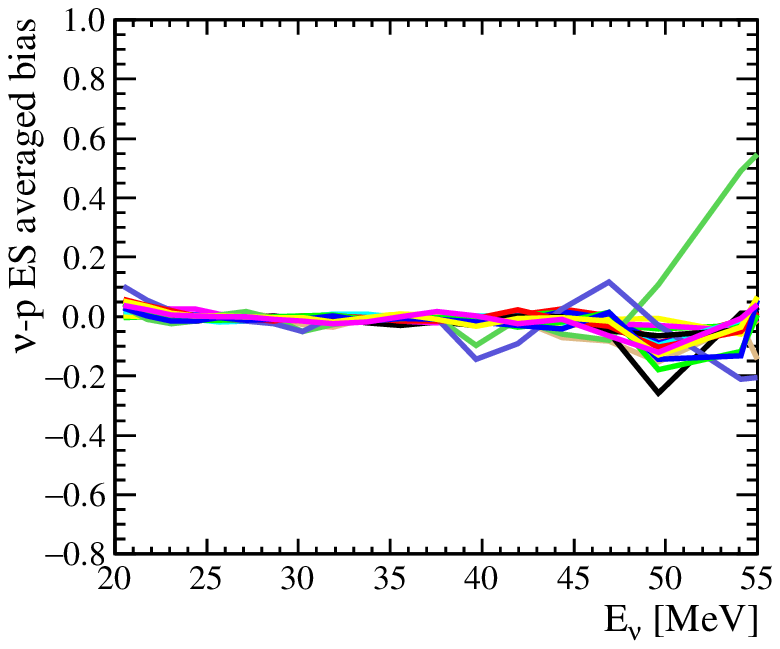}
\hspace{-0.8cm}
\includegraphics[width=0.54\textwidth]{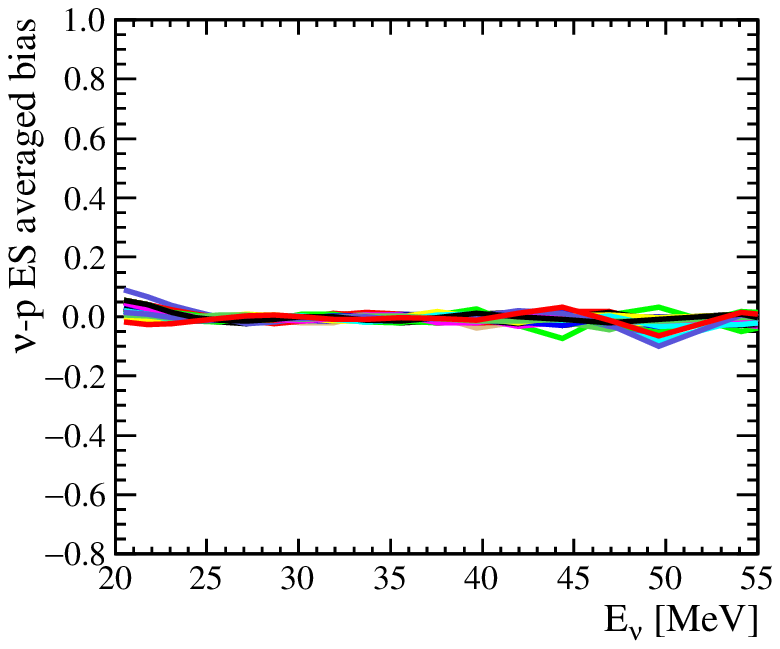}
\\
\vspace{-0.4cm}
\includegraphics[width=0.54\textwidth]{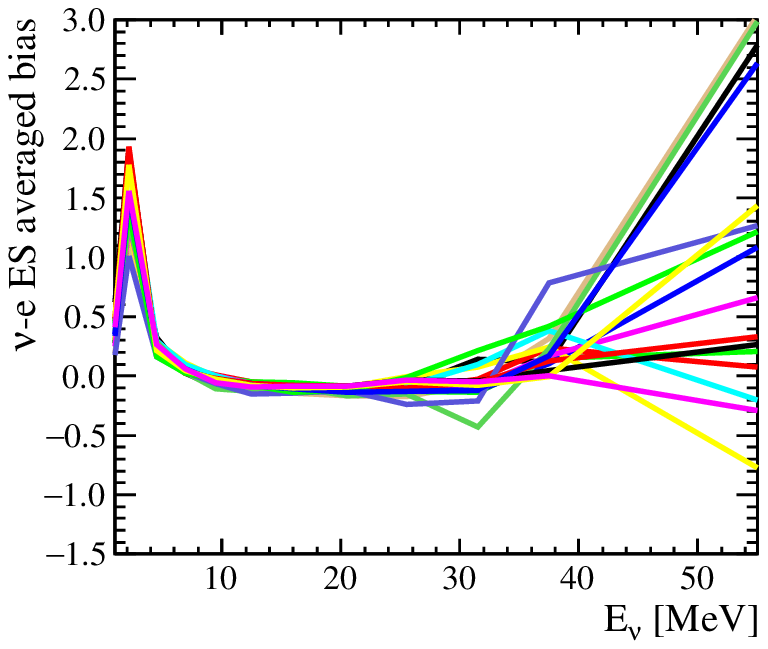}
\hspace{-0.8cm}
\includegraphics[width=0.54\textwidth]{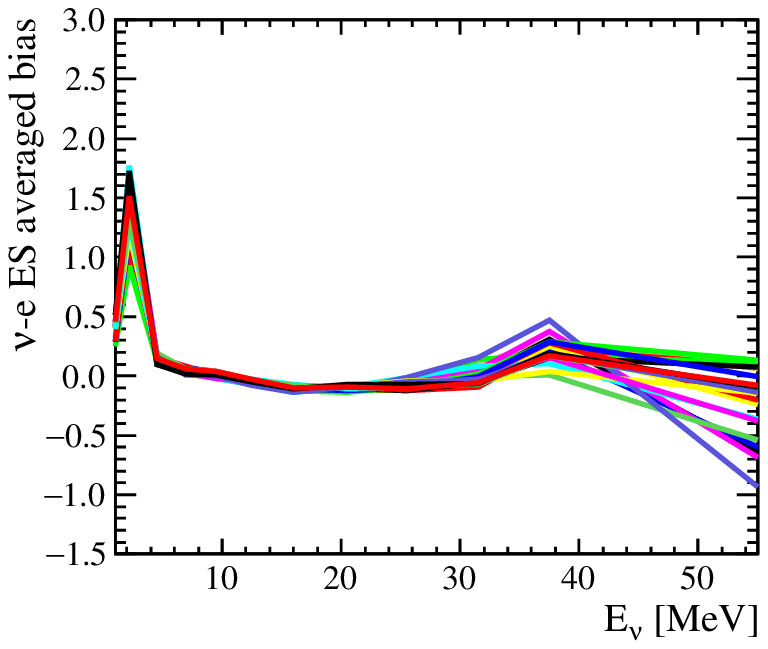}
\end{tabular}
\end{center}
\vspace{-0.1cm} \caption{The averaged bias of different numerical SN models as the function of the neutrino energy for the IBD (upper), $p$ES (middle) and $e$ES (lower) channels, where the SN distance of 10 kpc is assumed. The left column shows the results of 16 numerical SN models from the Garching group, while the right column for 21 SN models from the Japan group.
\label{fig:bias}}
\end{figure}

In the previous discussions, the KRJ parametrization of SN neutrino fluences has been used for numerical simulations. In order to demonstrate the robustness of the SVD unfolding approach, we now consider the SN neutrino fluences directly from the numerical SN models, which have been provided by the Garching~\cite{Hudepohl:2013} and the Japan~\cite{Nakazato:2013} groups. In total, 16 numerical models from the Garching group and 21 models from the Japan group are considered, covering a variety of progenitor star masses and different treatment of neutrino transport. For each numerical SN model we simulate 500 trials of the neutrino data and then apply the same reconstruction algorithm. For each trial, the bias is defined as the relative difference between the reconstructed and true neutrino spectra in each energy bin, which includes both the statistical fluctuation and the actual bias from the unfolding method. To reduce the statistical fluctuation, we further define the averaged bias among 500 trials as:
\begin{eqnarray}
{\rm Averaged\;Bias} = \frac{1}{N}\sum^{N}_{k=1} \left[\frac{n^{\rm true}_{k}(E^{i}_{\nu}) - n^{\rm reconstructed}_{k}(E^{i}_{\nu})}{n^{\rm true}_{k}(E^{i}_{\nu})}\right] \; ,
%     (17)
\end{eqnarray}
where $k$ is the index for the trial and $N=500$ is the number of total trials for one numerical model. This definition is applicable to all the IBD, $p$ES and $e$ES channels.

In Fig.~\ref{fig:bias}, the averaged bias of different numerical SN models for a SN distance of 10 kpc has been given for the reconstruction of SN $\overline{\nu}^{}_e$, $\nu^{}_x$ and $\nu^{}_e$ from the IBD, $p$ES and $e$ES channels, respectively. The left column is for 16 Garching numerical SN models, while the right column for the 21 Japan numerical SN models. From these plots, we can observe that the central parts of the curves are consistent with zero and thus no visible biases appear when the neutrino event numbers are large enough. Significant deviations from zero show up at one or two ends where the event statistics is limited. Further optimization on the binning schemes and unfolding regularization might be required to improve the performance. Comparing the left and right columns, one can see larger variations for the numerical SN models from the Garching group than those from the Japan group. This might be due to the fact that the Garching models cover an even wider range of core-collapse SNe, for which the basic properties and the main features of neutrino fluences differ greatly. Finally we conclude that the SVD unfolding method is a useful and robust method to reconstruct the neutrino energy spectra both in the IBD, $p$ES and $e$ES channels and for the three flavor neutrinos, at least for a particular region of neutrino energies.

Before closing this section, we make some remarks on the flavor conversions of SN neutrinos, which have so far been ignored in our discussions. Both the ordinary MSW matter effects~\cite{Wolfenstein:1977ue,Mikheev:1986gs} and the collective oscillations~\cite{Pantaleone:1992eq, Samuel:1993uw, Duan:2005cp, Duan:2006an, Hannestad:2006nj, Raffelt:2007yz,Duan:2009cd, Duan:2010bg, Chakraborty:2016yeg} lead to significant flavor conversions when neutrinos are propagating from the neutrino sphere to the surface of the SN. If only the MSW effects are taken into account, which become relevant in the SN mantle, there will be a partial or complete conversion between $\overline{\nu}^{}_e$ and $\nu^{}_x$ (and between ${\nu}^{}_e$ and $\nu^{}_x$), depending on the neutrino mass ordering. Therefore, the final energy spectrum of a given neutrino flavor would be the mixture of the initial spectra of two neutrino flavors with distinct average energies. On the other hand, the collective neutrino oscillations, which are induced by the coherent forward neutrino-neutrino scattering, may result in spectral splits of SN neutrino spectra~\cite{Duan:2010bg}. However, it remains an open question whether the collective neutrino oscillations do happen in the real SN environment~\cite{Chakraborty:2016yeg}. 

The SVD unfolding method under discussions is independent of any specific parametrization of the neutrino fluxes as well as the spectra at the detector, no matter whether neutrinos flavor conversions are included. As we have shown before, the event statistics is one of the most important factors that affect the reconstruction performance. The reconstruction tests with different numerical simulation models from the Garching and Japan groups demonstrate the robustness of the unfolding method for different inputs of the neutrino fluxes and energy spectra. On the other hand, the reconstructed neutrino spectra of three flavor neutrinos at the detector would be very useful to test the paradigm of neutrino flavor conversions, which will be presented in a separated work in the future.

\section{Summary}

For a future galactic core-collapse SN, a high-statistics measurement of neutrino signals can definitely be achieved. In particular, SN neutrinos of all three flavors $\overline{\nu}^{}_e$, $\nu^{}_e$ and $\nu^{}_x$ can be detected in the IBD, $e$ES and $p$ES channels in a single large LS detector, such as JUNO. The precise determinations of SN neutrino spectra and flavor contents will be crucially important to verify the delayed neutrino-driven explosion mechanism, and to reveal the complicated pattern of SN neutrino flavor conversions~\cite{Mirizzi:2015eza}.

In this paper, we apply the SVD approach with a proper regularization~\cite{Hocker:1995kb} to reconstruct SN neutrino energy spectra for all flavors in a $20~{\rm kiloton}$ LS detector with the resolution similar to JUNO. First of all, we explain how to calculate the SN neutrino events, and introduce an analytical recipe to reconstruct neutrino spectra in the ideal case where the finite energy resolution is neglected. Then, taking account of the realistic energy resolution, we simulate the neutrino data by using the parametrized neutrino spectrum (e.g., the KRJ parametrization with $\langle E^{}_{\nu^{}_e}\rangle = 12~{\rm MeV}$, $\langle E^{}_{\overline{\nu}^{}_e}\rangle = 14~{\rm MeV}$ and and $\langle E^{}_{\nu^{}_x}\rangle = 16~{\rm MeV}$). Starting with the reconstructed spectra, we also combine the IBD, $e$ES and $p$ES channels to extract the original neutrino spectra in a common region of neutrino energies. It turns out that for a SN at 10 kpc there is no problem to accurately reconstruct $\overline{\nu}^{}_e$ and $\nu^{}_x$ spectra, but the reconstruction of $\nu^{}_e$ is limited by the relatively low statistics. Finally, instead of the parametrized SN fluences, the numerical SN models from the Garching and Japan groups are analyzed as well. In addition, the robustness and validity of the SVD approach for both analytical and numerical neutrino data are demonstrated to be excellent.

Even though a single large LS detector like JUNO has the great potential to detect SN neutrinos of all three flavors, the statistics in the $e$ES channel is quite low for a SN at 10 kpc. In this case, it is very important to notice that the WC detectors (e.g., Hyper-Kamiokande) and the LArTPC detectors (e.g., DUNE) are exceptionally good at observing the SN $\nu^{}_e$. Therefore, different types of future large detectors are actually complementary to each other. A combined analysis of the LS, WC and LArTPC detectors in reconstructing SN neutrino spectra will be very interesting and deserve dedicated studies in the near future.

\section*{Acknowledgements}

The authors thank Feng-Peng An, John Beacom and Basudeb Dasgupta for valuable discussions and helpful suggestions, and to the Garching group for providing neutrino data from their numerical simulations. This work was in part supported by the National Natural Science Foundation of China (NSFC) under Grant No.~11775232, by the Strategic Priority Research Program of the Chinese Academy of Sciences under Grant No.~XDA10010100, by a grant from the Ministry of Science and Technology of China under Grant No.~2013CB834303, by the Key Laboratory of Particle Physics and Particle Irradiation (Shandong University) of the Ministry of Education, by the National Recruitment Program for Young Professionals and the CAS Center for Excellence in Particle Physics (CCEPP).

%by the joint project of NSFC under Grant no.~11611530683 and the Russian Foundation for Basic Research under Grant no. 17-52-53133 ${\rm GFENa}$,

\end{document}